\documentclass[fleqn,10pt]{wlscirep}
\usepackage[utf8]{inputenc}
\usepackage[T1]{fontenc}
\usepackage{ulem}

\usepackage{booktabs}
\newcommand{\ra}[1]{\renewcommand{\arraystretch}{#1}}

\title{A quantitative description of the transition between intuitive altruism and rational deliberation in iterated Prisoner’s Dilemma experiments}

\author[1,2,3,*]{Riccardo Gallotti}
\author[4,5]{Jelena Gruji\'c}

\affil[1]{Instituto de F\'{\i}sica Interdisciplinar y Sistemas Complejos IFISC (CSIC-UIB), Palma de Mallorca, Spain.}
\affil[2]{Center for Complex Systems $\&$ Brain Sciences (CEMSC$^3$), Universidad Nacional de San Martin, Buenos Aires, Argentina.}
\affil[3]{Fondazione Bruno Kessler, Trento, Italy.}

\affil[4]{AI lab, Computer Science Department, Vrije Universiteit Brussel, Brussels, Belgium.}
\affil[5]{MLG, D\'epartement d'Informatique, Universit\'e Libre de Bruxelles, Brussels, Belgium.}
\affil[*]{To whom correspondence should be addressed; E-mail:  rgallotti@gmail.com.}

\begin{abstract} 
What is intuitive: pro-social or anti-social behaviour? To answer this fundamental question, recent studies analyse decision times in game theory experiments under the assumption that intuitive decisions are fast and that deliberation is slow. These analyses keep track of the average time taken to make decisions under different conditions. Lacking any knowledge of the underlying dynamics, such simplistic approach might however lead to erroneous interpretations.

Here we model the cognitive basis of strategic cooperative decision making using the Drift Diffusion Model to discern between deliberation and intuition and describe the evolution of the decision making in iterated Prisoner's Dilemma experiments. 

We find that, although initially people's intuitive decision is to cooperate, rational deliberation quickly becomes dominant over an initial intuitive bias towards cooperation, which is fostered by positive interactions as much as frustrated by a negative one. However, this initial pro-social tendency is resilient, as after a pause it resets to the same initial value. 

These results illustrate the new insight that can be achieved thanks to a quantitative modelling of human behavior.
\end{abstract}

\begin{document}

\flushbottom
\maketitle
%
%
\thispagestyle{empty}

\section*{Introduction}

Decision times have emerged as a new important aspect in experimental game theory. They have been measured in a wide range of games like: Ultimatum games ~\cite{branas2017strategic, cappelletti2011being}, Modified Dictator games~\cite{piovesan2009fast} and Public Goods games~\cite{rand2012spontaneous,lotito2013cooperation,tinghog2013intuition,evans2014reaction,rand2014social}. 
Some studies, based on the premise that decisions which take less time are more intuitive, suggest that making unselfish, cooperative decisions is a human instinct that is then undermined by rational deliberation~\cite{rand2012spontaneous,rand2014social}. 
The claim that showed that people under pressure make more cooperative decisions in Public Goods Games~\cite{rand2012spontaneous} failed a recent Registered Replication Report~\cite{bouwmeester2017registered} including 21 independent experiments.
Other criticisms include that, after controlling for the strength-of-preference between the two options (as the players could have strong preference of one choice over the other), there is no significant difference between the pro-social and the selfish behavior~\cite{krajbich2015rethinking}. A more detailed meta-analysis of a large number of experiments suggests that things are more complicated, claiming that deliberation inhibits what they called `pure cooperation', however it does not appear to inhibit the strategic cooperation~\cite{rand2016cooperation}. Studies using fMRI scanners to monitor brain activity of human subjects participating in Game Theory experiments show that selfish participants will cooperate when they have incentive to cooperate: consequently cooperation can also be a result of longer deliberation, not just intuitive acts of pro-social individuals~\cite{emonds2011comparing,lambert2017trust}.  

\begin{figure*}[ht!]
\centerline{
\includegraphics[angle=0,width=0.48\textwidth]{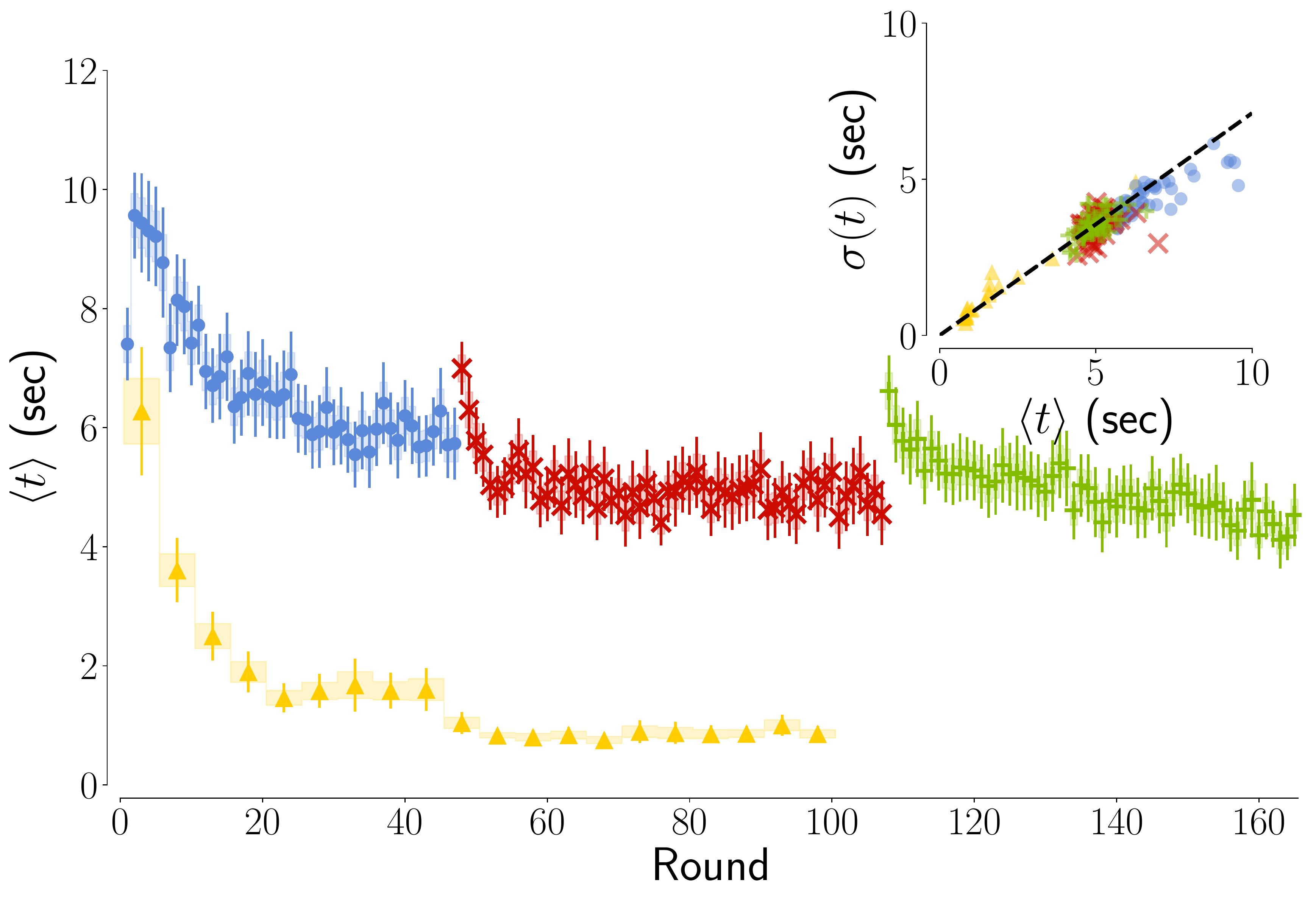}
\qquad
\includegraphics[angle=0,width=0.48\textwidth]{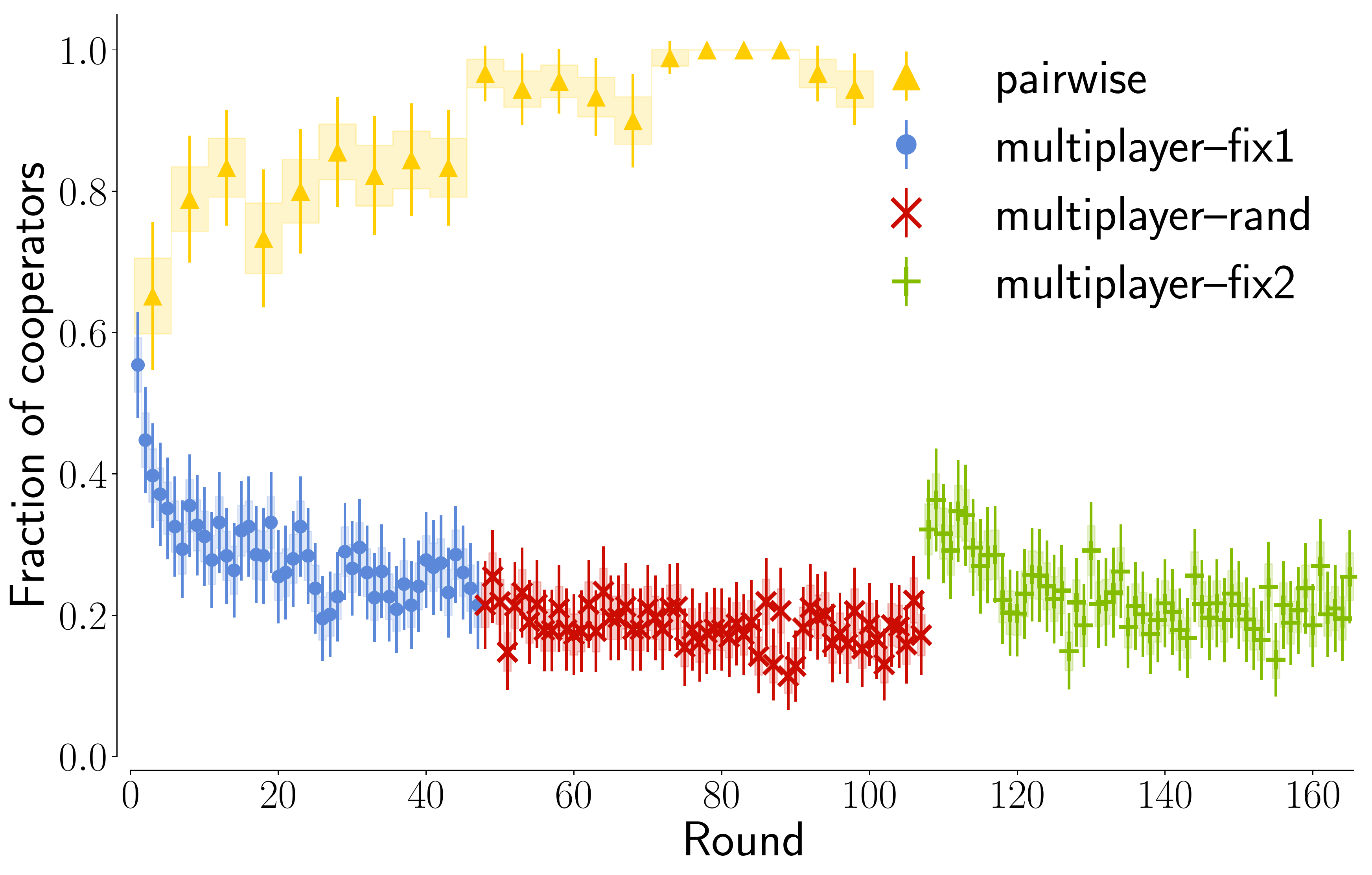}
}
\caption{
{\bf A classical view on cooperation level and decision times in our iterated Prisoner's Dilemma experiments}. 
{\bf (Left)} The decision times start from a similar initial value of $\approx 7$ seconds, and then follow two different behaviours in the two experiments we analyze. In the inset we illustrate the direct proportionality between mean and standard deviations. 
{\bf (Right)} The cooperation level also starts from a similar initial value of $\approx 0.6$ for the two experiments. In the pairwise experiment full cooperation is attained, while in the multiplayer experiment the majority of subject opt for defection. 
In both panels and in the inset blue circles, green squares, and orange diamonds represent the three phases of the multiplayer experiment (the distinction between these phases is not important at the moment and it will be explained later in the text) while red hexagons the pairwise experiment. In this and in all figures of this paper, the error bars represent the very broad 95\% confidence interval associated to 1.96 standard deviations, while the shaded area represent one standard deviation.}
\label{figure1}
\end{figure*}

The standard procedure in these papers is to perform game theory experiments with human subjects and to track the mean decision times (RT) for pro-social and selfish decisions. It was assumed that the players think less to make intuitive decisions an longer to make deliberate decisions, and therefore concluded that the type of decisions with smaller RTs are intuitive and those with the longer RTs are deliberate. This simplified approach might be the reason why, thus far, a consensus is far from being reached. There are a number of reasons why tracking just the mean RT values is not a good idea. First, the experimental RT distributions have a number of properties which prevent it from being properly described by only mean value and standard deviation (as we do in Fig.~\ref{figure1} left, where we describe the average RTs for the two iterated Prisoner Dilemma experiments analyzed in this paper, one multiplayer played with 8 people at the same time and one played with only one other person). For example the distribution it is not Gaussian but a skewed function whose skew actually increases with the task difficulty, while the mean value and standard deviation are proportional to one another~\cite{forstmann2016sequential} (see also the inset in Fig.~\ref{figure1} left). Second, a number of factors influence the speed of decision making, including the quality of information processing (how fast one accumulates information) or response caution (how careful we are not to make errors). An a-priori bias, telling us what the intuitive decision is, is just one of the factors that could make the RT short. Therefore, before we can talk about intuitive and deliberate decisions, we need first to identify a model allowing us to disentangle the different factors of influence.

\subsection*{Drift Diffusion Model}

Interestingly, in cognitive science and neuroscience, a number of theoretical models have emerged that can be used to explain the decision making processes.
The most prominent one among these is the Drift Diffusion Model (DDM)~\cite{Bogacz:2006fj}, which assumes a one-dimensional random walk behavior representing the accumulation in our brain of noisy evidence in favour of two alternative options~\cite{Smith:2004jo}.

DDM connects neuroscience to economic behavior by approximating the dynamics of the neurobiological process in act while the decision is made~\cite{Glimcher:2004et}. It thus represents a view complementary to other models describing the subject actions'. It focus on the shorter timescale~\cite{sapolsky2017behave} of neurobiology rather than attempting at finding the cause of observed behavior as a consequence of natural selection (e.g. evolutionary dynamics), learning process (e.g. operant conditioning), or of the individuals' strategies (e.g. conditional cooperation \cite{fischbacher2001people}, which important role in the multilayer experiment studied in this paper has already been described~\cite{grujic2010social}). It has in particular the unique feature of precisely outlining the statistics of decision times, while it is intrinsically limited in its ability of identifying the causes of single actions taken.

\begin{figure*}[ht!]
\centerline{
\includegraphics[angle=0,width=0.48\textwidth]{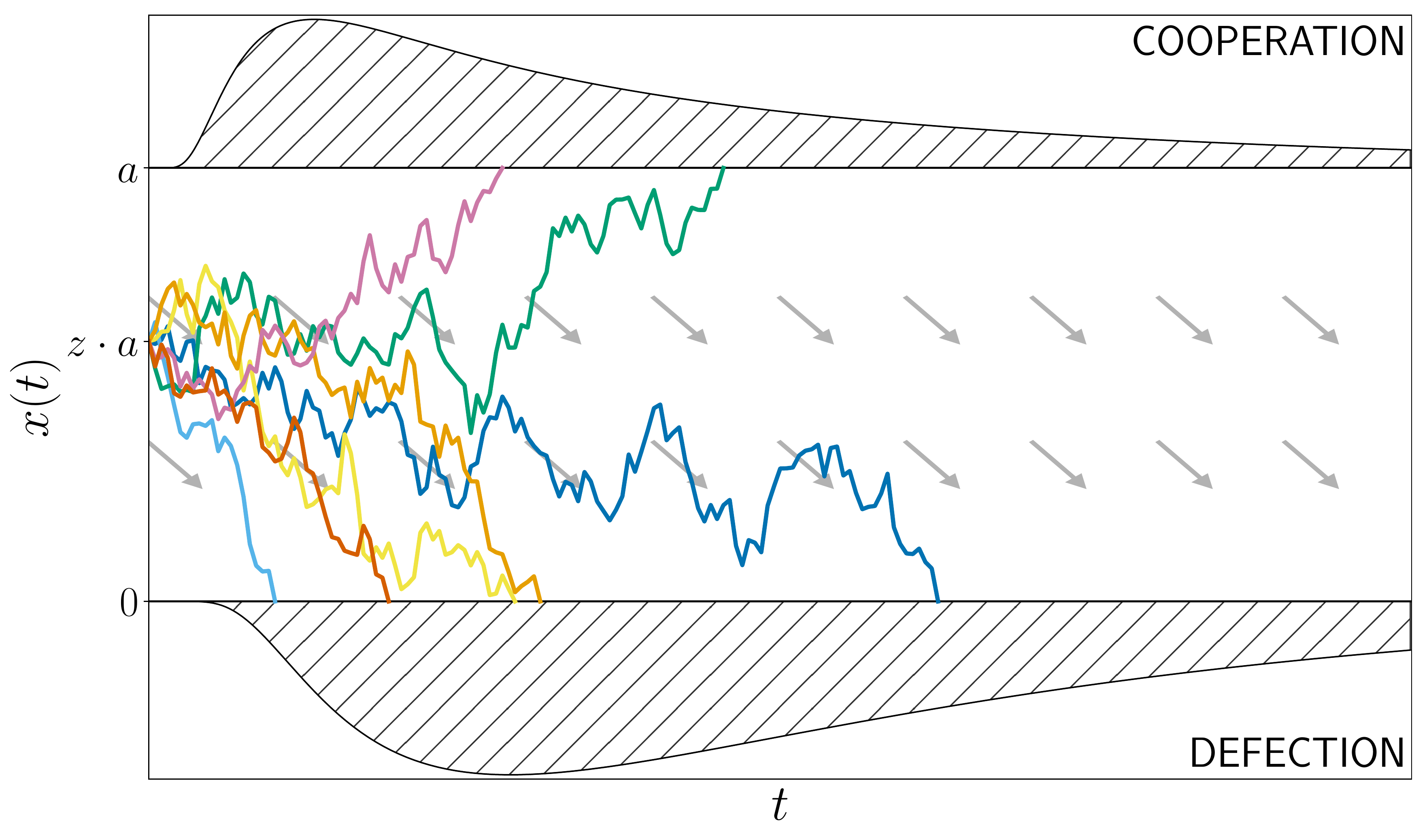}
\qquad
\includegraphics[angle=0,width=0.47\textwidth]{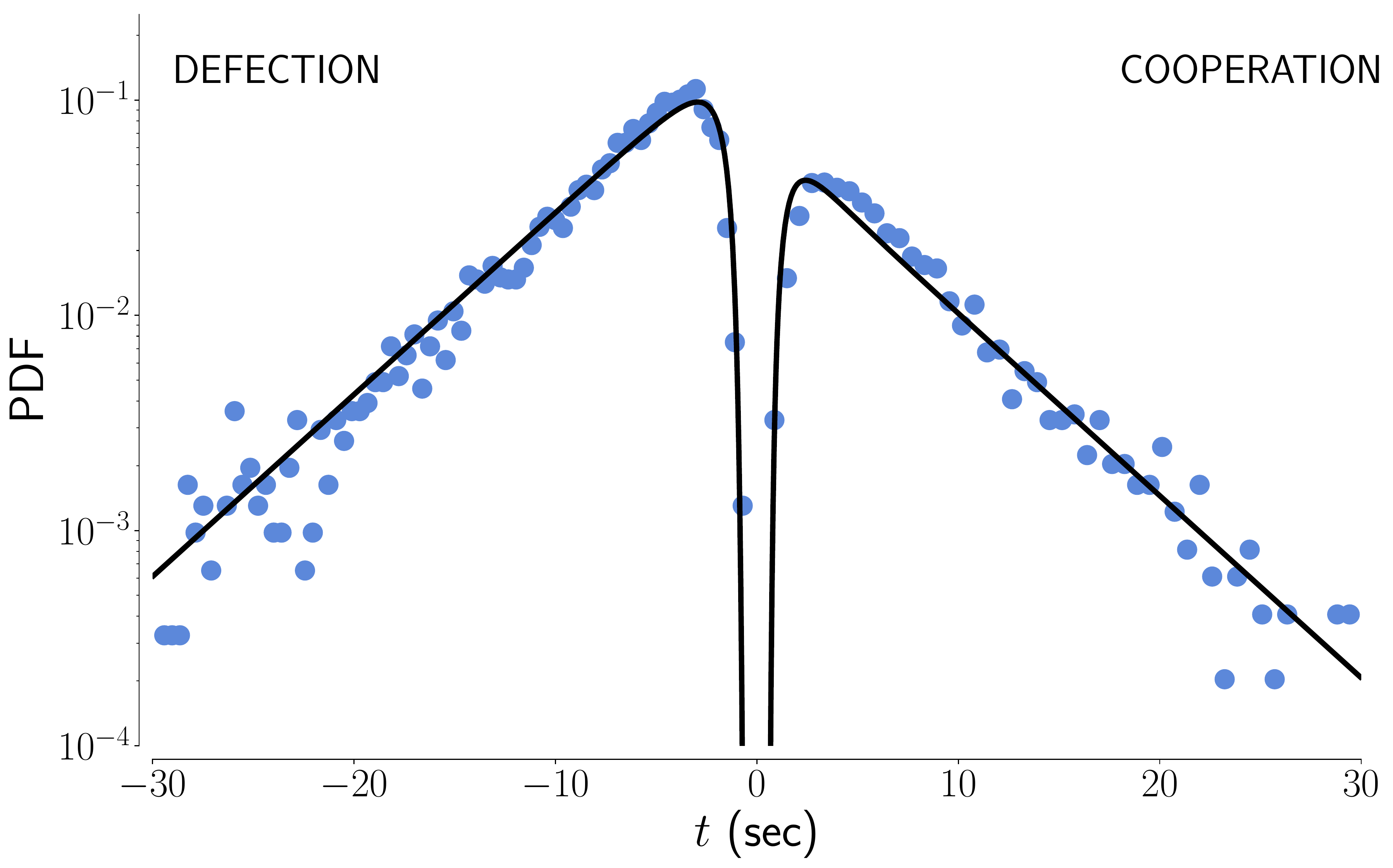}
}
\caption{
{\bf Drift Diffusion modelling for decision times and cooperation levels.} 
{\bf (Left)} An illustration of the DDM: starting from an initial condition $z\cdot a$, the agents accumulate random evidence in favour of one of two alternative decisions. The $x=a$ threshold is associated to cooperation and the $x=0$ threshold to defection. Once the amount of evidence reaches one of the thresholds, the associated decision is made. The arrows indicate the presence of a negative drift towards defection, as we observe in the multiplayer experiment. The two curves with shaded area represent the two parts of the probability distribution for the response times, one for the cooperation the other for defection, which are expected to differ in both shape and area.
{\bf (Right)} The experimental distribution for fix1 phase, fitted with the theoretical curves for the DDM ($r^2 = 0.97$). To distinguish, the response times of defection and cooperation we display them separately (reaction times for defection on the negative axis and those for cooperation on the positive axis). The total areas  under the two curves is normalised to one, so that the area under each of the two curves represents the proportion of defectors and cooperators. The logarithmic scale highlights the short tail of the distribution.
}
\label{figure2}
\end{figure*}

More in detail, in the DDM at each moment subjects randomly collect evidence in favour of one of two alternative choices, which are in our case cooperation and defection. The continuous integration of evidence in time is described by the evolution of an one-dimensional brownian motion (see Methods), whose stochastic character is a consequence of the noisy nature of the evidence~\cite{Busemeyer:1993bm,Brunton:2013kg}. The process is starting from a possibly biased initial condition, and the two options are associated to two absorbing barriers (see Fig.~\ref{figure2} left). The distribution of first passage times at those thresholds has been successfully used to model decision time in a wide range of contexts~\cite{forstmann2016sequential}. The most typical context is visual decision making~\cite{Glimcher:2003cy} where a subject (a human or another primate) needs to determine as quickly as possible the direction of a cloud of dots~\cite{ratcliff2008diffusion}. Comparing empirical RTs with the model allows us to evaluate its free parameters (see Table~\ref{table_parameters}):  i) the threshold $a$, is the quantity that quantifies response caution; ii) the drift rate $v$ is a measure of subjects' ability to gather evidence, also dependent on task difficulty; iii) the bias $z$ represents the a-priori inclination for one of the alternatives (with $z=0.5$ representing the unbiased scenario, see Methods). We also introduce a fourth parameter, the non-decision time $t_0$, that accounts here for the perceptual and motor processes associated with the task, processes which play a very minor role here given the longer characteristic timescale of the RTs in our experiments. 

In absence of bias or drift, decisions are completely random and the cooperation rate is expected to be $C^{teo}(v=0,a,z=0.5) = 0.5$. Using the DDM, the deviation of the experimental cooperation rate $C$ from random behavior can be decomposed into a contribution due to $v$ (rationality) and a contribution due to $z$ (intuition). If only the bias is absent, one would expect a cooperation rate $C^{teo}(v,a,z=0.5)$. Therefore, we identify as contribution of the rational deliberation the difference $\Delta C_{rat} = C^{teo}(v,a,z=0.5)-0.5$, and as contribution of the intuitive bias the difference between the empirical value and what expected without bias $\Delta C_{int} = C-C^{teo}(v,a,z=0.5)$. 

For quantifying how much a decision is influenced by the a-priori bias, in this paper we introduce the `rationality ratio'

\begin{equation}
R = |\Delta C_{rat}|/(|\Delta C_{rat}|+|\Delta C_{int}|) \ .
\label{eq_R}
\end{equation}

\begin{table*}[h]
\begin{center}
\ra{1.3}
\begin{tabular*}{\textwidth}{@{\extracolsep{\fill}}lclc@{}}
\toprule[1pt]
DDM parameter  & Symbol & Interpretation  & Used in fits\\
\hline
threshold			& $a$					& information required -- response caution (perceived difficulty)				            & yes			\\
drift rate			& $v$					& information gathering -- task complexity		& yes			\\
bias				& $z$					& a-priori inclinations					        & yes			\\
non-decision time	& $t_0$					& perceptual and motor process			        & yes		    \\
\bottomrule
\end{tabular*}
\end{center}
\caption{
{\bf Free parameters of the model.} 
\textnormal{ Across all analysis done in this paper, we considered all four parameters of the DDM as free and are estimated directly from a fit to the subset of data considered. Note that high values of the drift rate $v$ represent fast information gathering which in turn can be sometimes associated to lower task complexity.}}
\label{table_parameters}
\end{table*}

\subsection*{Experiments}

The application of DDM to more deliberate economic decision making has been uncertain until recently~\cite{forstmann2016sequential}, and the model is still largely unknown in the experimental game theory community. 
In the last year, some experiments have showed that DDM can be applied to economic experiments, however only in situations when the player has the full information. For example when the outcome does not depend on the action of the opponents such as in the Dictators Game~\cite{krajbich2015common,hutcherson2015neurocomputational,chen2018biased}, or when the action of the opponent is already known, such as on the receivers side of the Ultimatum Game~\cite{krajbich2015common}. The decision making in the aforementioned works is value-based, but not strategic. The players know exactly how much will be their payoff for each decision, and this payoff does not depend at all on the actions of other players.

Here we extend the use of the DDM to outline the cognitive basis of cooperative decision making and characterise the evolution of a subject's behavior when facing strategic choices in game theory experiments, where the decision is dependent on the unknown action of the opponent. 
The application of DDM to Prisoner's Dilemma games is not obvious given the current state of the art, since in the deliberation process includes forming an opinion on what the other player would do. For example, the model proposed in~\cite{hutcherson2015neurocomputational} takes into account how much the players care about themselves and about other players. A similar approach to decisions in complex strategy games would require the development of a non-trivial addition to the model trying to predict the decision of the other player, but there is no obvious way to include this in the model.

In particular, we examine the results of two different weak Prisoner's dilemma experiments iterated over a large number of rounds~\cite{grujic2010social,grujic2012three,grujic2014comparative}: 
i) a pairwise game with 16 players associated in fixed couples, iterated 100 times;
ii) a multiplayer game on 13x13 lattice with 8 neighbours, where in every round each player makes a single action (cooperate or defect) which applies to a game against each neighbour. All participants of the multiplayer experiment played a total of 165 rounds in 3 separate phases: fix1, rand, fix2 (in that order). In the phases fix1 and fix2 the network is fixed and all players play with the same neighbours. In the rand phase the network is randomly shuffled after every round, so the players always play with different neighbours. (For more details see the Methods or the original papers~\cite{grujic2010social,grujic2012three,grujic2014comparative}). The alternation between a static and a changing 
network was originally intended as a sort of ``control'' in  the experimental setup. Serendipitously, it allowed us to compare a more complex task (the static network, where the history of all interactions with the different neighbors can be in principle considered) with an easier one (the random network, where there is no information on the previous actions by other players that one would need to consider in order to make the decision.

In both experiments, the cooperation level starts from a value close to 60\%. In the multiplayer game this then converges to low cooperation ($\approx20\%$), while in the pairwise game an almost full cooperation is attained after about 70 rounds (see Fig.~\ref{figure1} right). Thus, these two experiments allow us to observe two different scenarios: one where the cooperation is not established, and one where it is. Thanks to this broad variability in subjects' behavior, we can also control if the strength of preference is influencing or not our conclusions~\cite{krajbich2015rethinking}.

It might be surprising that we chose two experiments with such different setups. However, we have a good reason for that. We originally only analysed the large network experiment, but eventually we realized that we need to test if our conclusions would still hold in cooperative environment. Achieving cooperation in the large experiment by tweaking small number of parameters was very uncertain. Furthermore, this kind of experiments where almost 200 people play the game at the same time are logistically very difficult and quite expensive. Finally, it was unnecessary to have a similar experiment, when the only thing we wanted to know is how the cooperative environment influence the bias and the drift. Therefore we opted to perform the simplest experiment where we knew the cooperation will be established. Evidently, since the details of the two experiment are quite different, in order to avoid drawing any conclusion that could be influenced by these differences here we only discuss the different cooperation level finally attained.

\section*{Results}


In Figures~\ref{figure2} right and Supplementary Fig. 3 we show the Probability Density Functions (PDF) of the Response Times $t$ where we separated reaction times for defections and cooperations by assigning negative reaction times values to the defections. The curve is normalized to one considering both positive and negative values, therefore the larger area under the negative curve, as compared to the positive curve, corresponds to a larger number of defections in the experiment. The comparison between the empirical scattered data and the theoretical curves (solid lines) shows that DDM successfully fits the empirical RTs of the different phases of our experiment. In the following, we show what one can observe by tracking the evolution of the DDM parameters in our experiments. This allows us in general to describe the learning process of the subjects of game theory experiments from a novel perspective aiming at
distinguishing between rational deliberation (described by the drift $v$,) and intuition (associated to the a-priori bias $z$). The fits have been performed using the HDDM python tool~\cite{wiecki2013hddm}, which fits simultaneously the distribution of decision times for cooperative choices, defections, and the fraction of cooperation using Hierarchical Bayesian Estimation. (In Supplementary Fig. 4) we provide the $R^2$ values of all the fits proposed in this paper.)

\begin{figure*}
\begin{center}
\begin{tabular}{cc}
\raisebox{2.5cm}{(a)} \includegraphics[angle=0, width=0.45\textwidth]{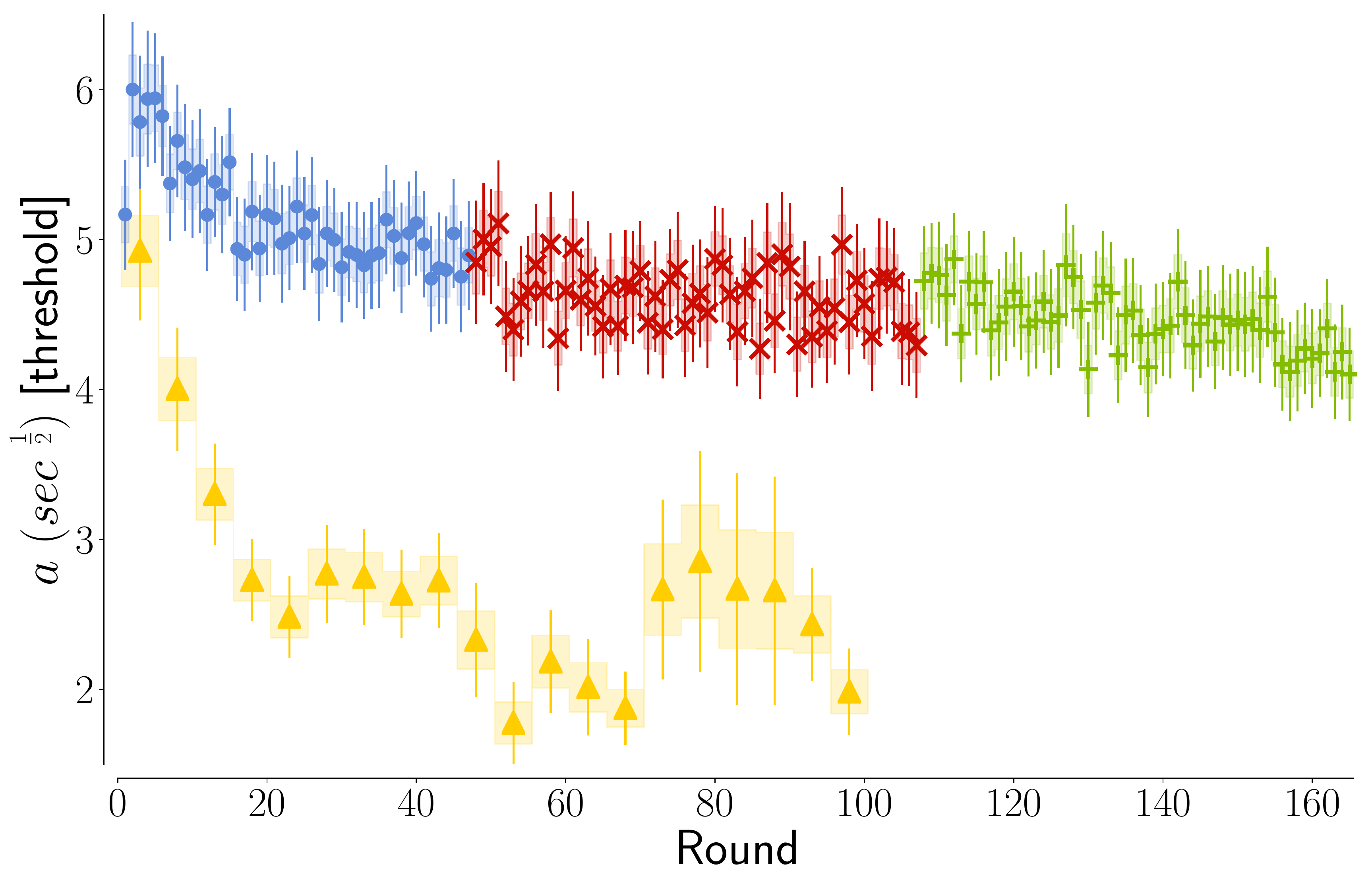}&
\raisebox{2.5cm}{(b)} \includegraphics[angle=0, width=0.45\textwidth]{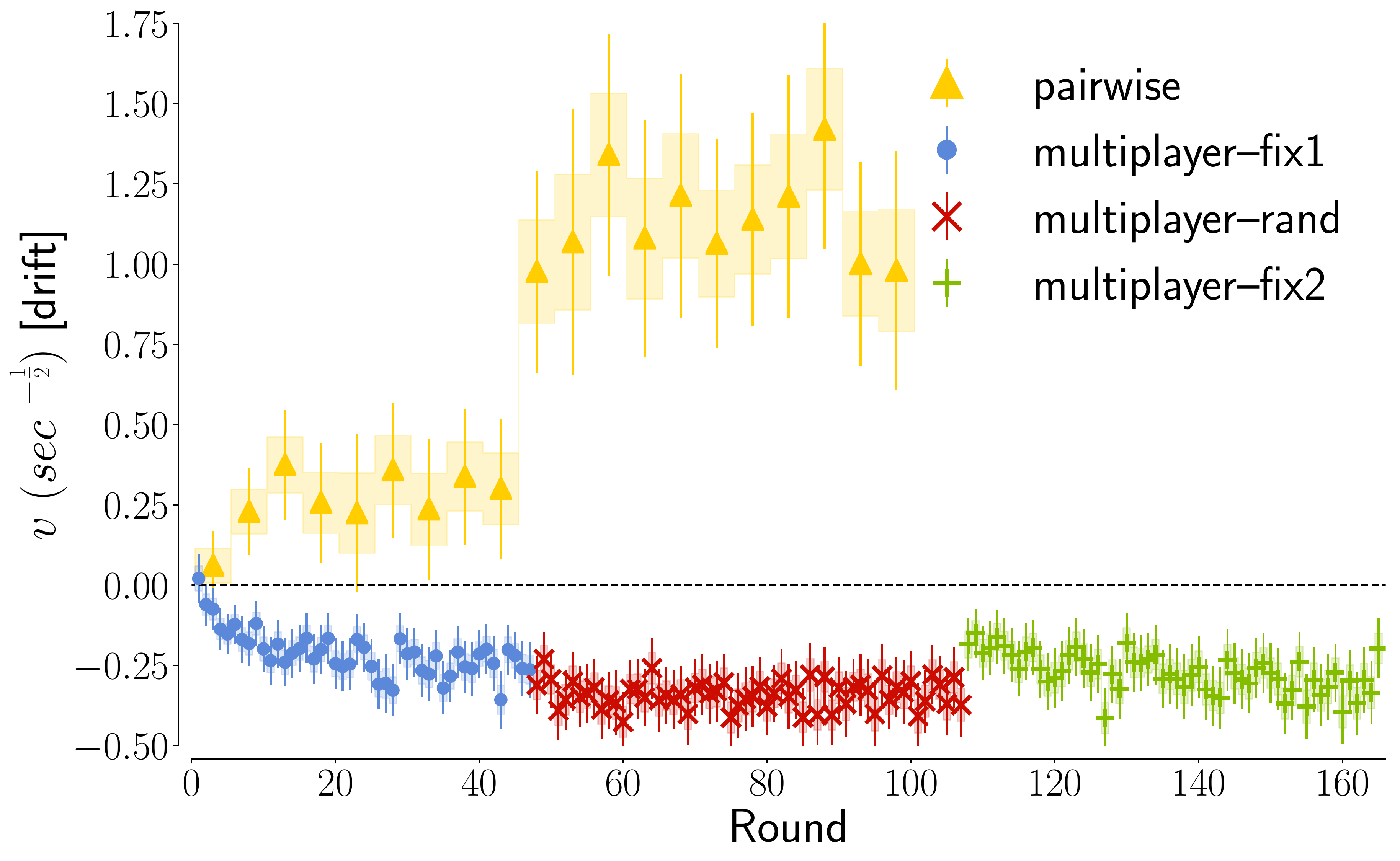} \\
\raisebox{2.5cm}{(c)} \includegraphics[angle=0, width=0.45\textwidth]{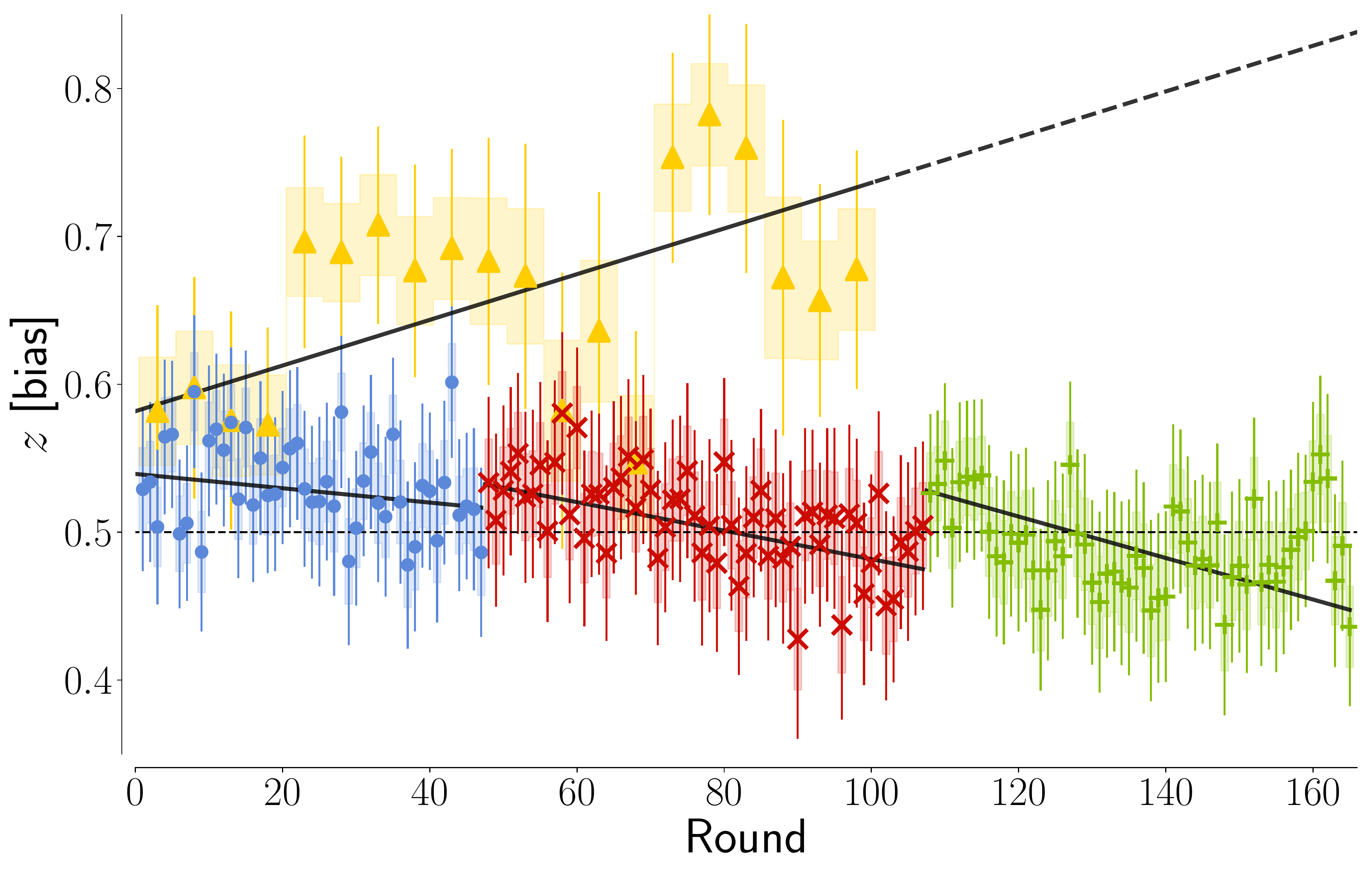} &
\raisebox{2.5cm}{(d)} \includegraphics[angle=0, width=0.45\textwidth]{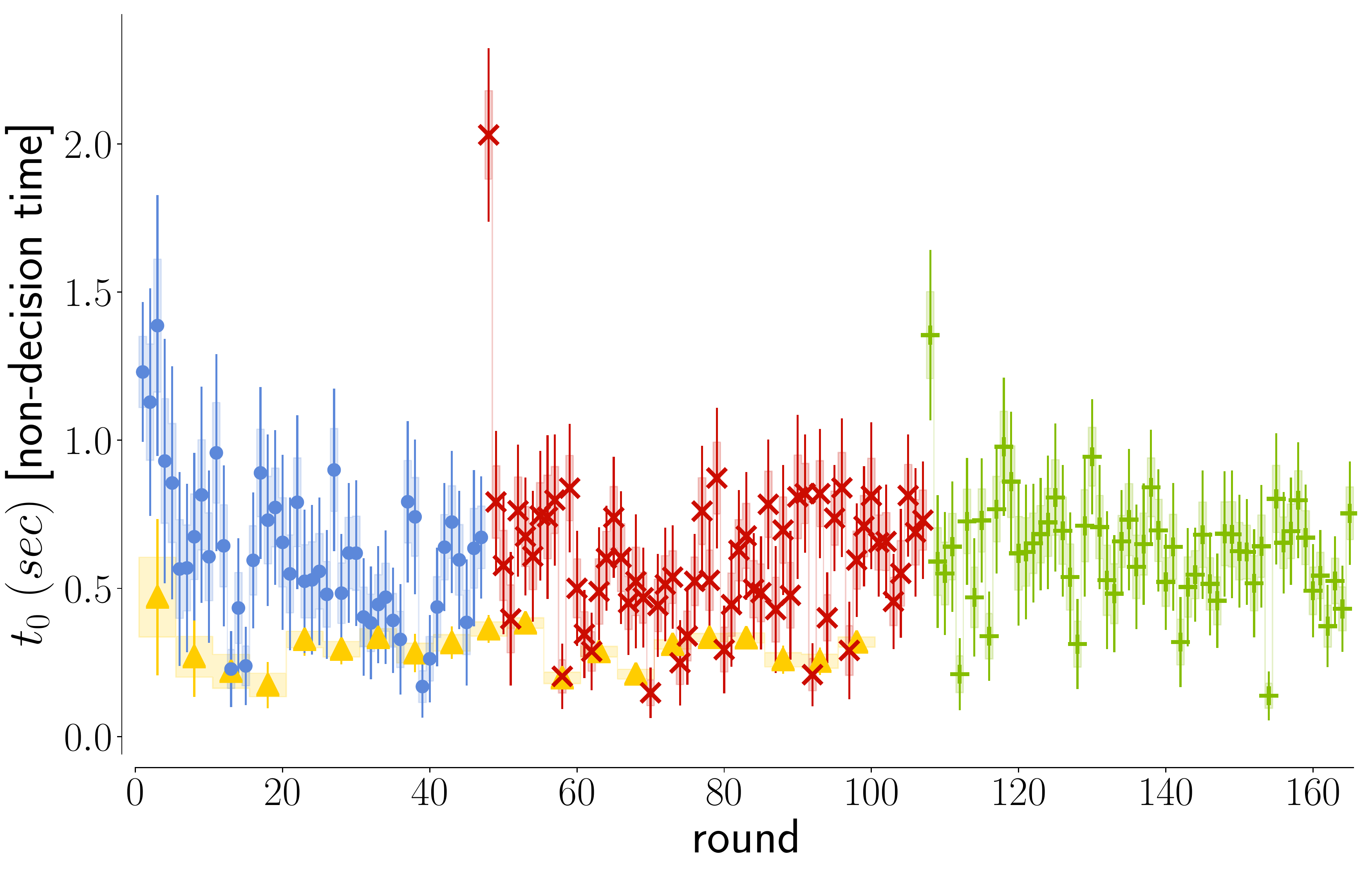} \\
\end{tabular}
\end{center}
\caption{
{\bf Evolution of the DDM parameters  in the two experiment.} 
{\bf (a)} In the first round, the threshold $a$ is $\approx 5$  sec$^{\frac{1}{2}}$ in both experiments. For the pairwise experiment we then observe a decreasing trend, while for the multiplayer experiment we observe an increase in the second round, followed by a progressive drop in the course of the experiments.
{\bf (b)} The absolute value of the drift speed $|v|$ also starts from a common value approximatively zero. Its absolute value progressively increases for both experiments, showing how players process the information faster. The sign differs between the two experiments, because for the multiplayer case the gathered information suggests to defect while for the pairwise interaction it suggests to cooperate. The random phase of the multiplayer experiment has higher $|v|$, which is consistent with the fact that the setup of the random phase is easier. 
{\bf (c)} Both experiment suggest an initial bias towards cooperation $z \approx 0.55$. The bias then changes progressively in the direction of the average behavior of the other participants: positive bias for the cooperation in the pairwise experiment, negative bias for the defection in the multiplayer experiment (in Supplementary Fig. 5 we show that this trend does not depend by the player's actions). In the multiplayer experiment, after each phase the bias resets to its initial value of 0.55, suggesting a resilience in the human bias towards cooperation.
{\bf (d)} The non decision time $t_0$ drops after a few rounds to a constant value of $\approx 0.6$ sec for the multiplayer experiment and $\approx 0.3$ sec for the pairwise experiment. At the first round of the second and third phases of the multiplayer experiment, we observe a clear outlier, possibly accounting for the fact that the individuals were not ready for the next phase.
(In this and all figures, the shaded areas represent the s.e.m and the error bars the 95\% confidence level.)
}
\label{figure3}
\end{figure*}


{\bf Threshold} -- corresponds to the perceived difficulty of the task and therefore drops over time in all experiments. In both experiments, the threshold starts from the same value of $a \approx 5$ sec$^\frac{1}{2}$, in the first round where no information on the others' behavior is available (see Fig.~\ref{figure3} (a))  Then, the threshold parameter drops for the pairwise experiment. While it initially grows for the more complex multiplayer experiment, a drop can be then observed starting from round 5.

\begin{figure*}[ht!]
\begin{tabular}{cc}
\raisebox{2.5cm}{(a)} \includegraphics[angle=0, width=0.45\textwidth]{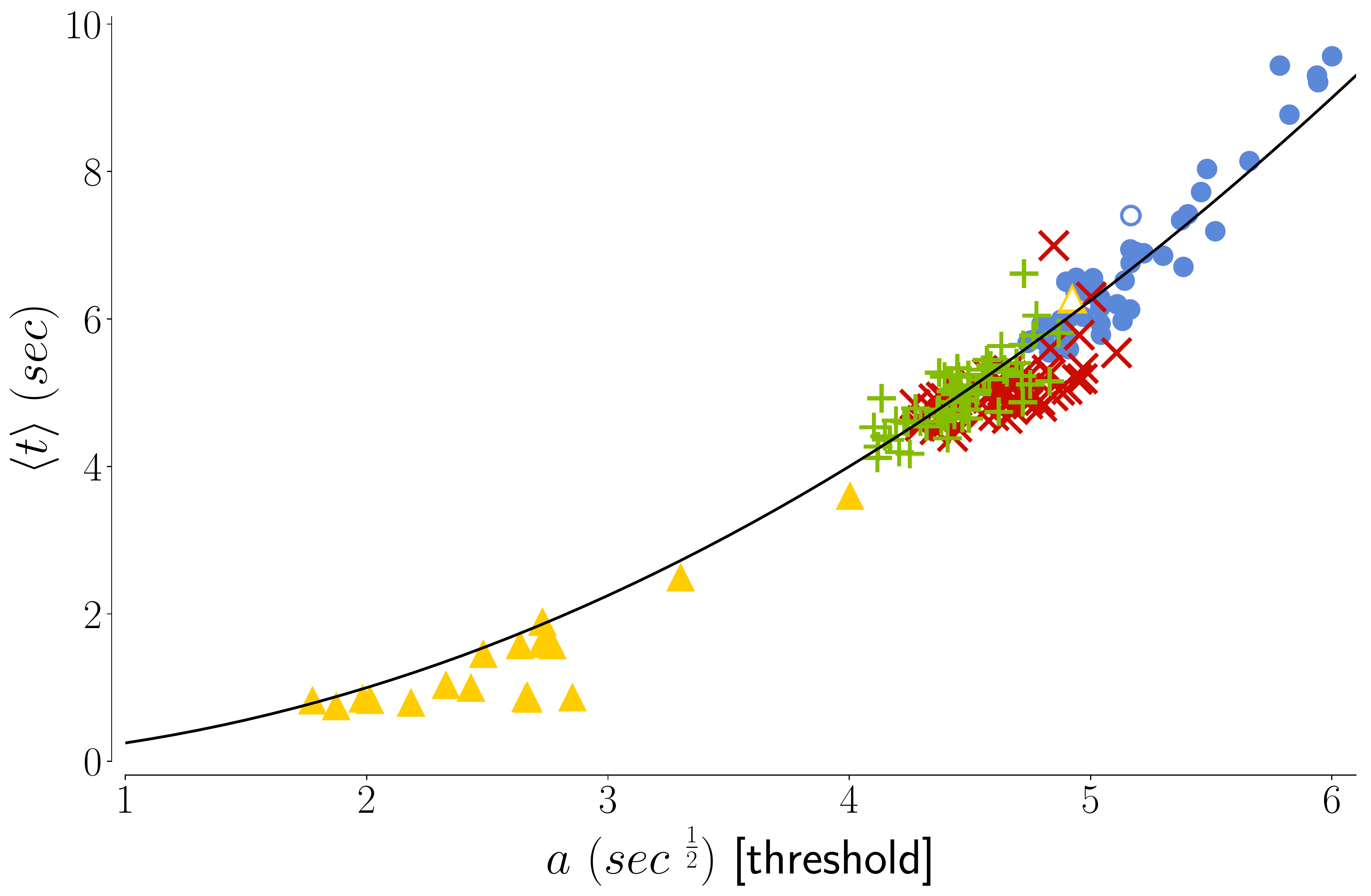}&
\raisebox{2.5cm}{(b)} \includegraphics[angle=0, width=0.45\textwidth]{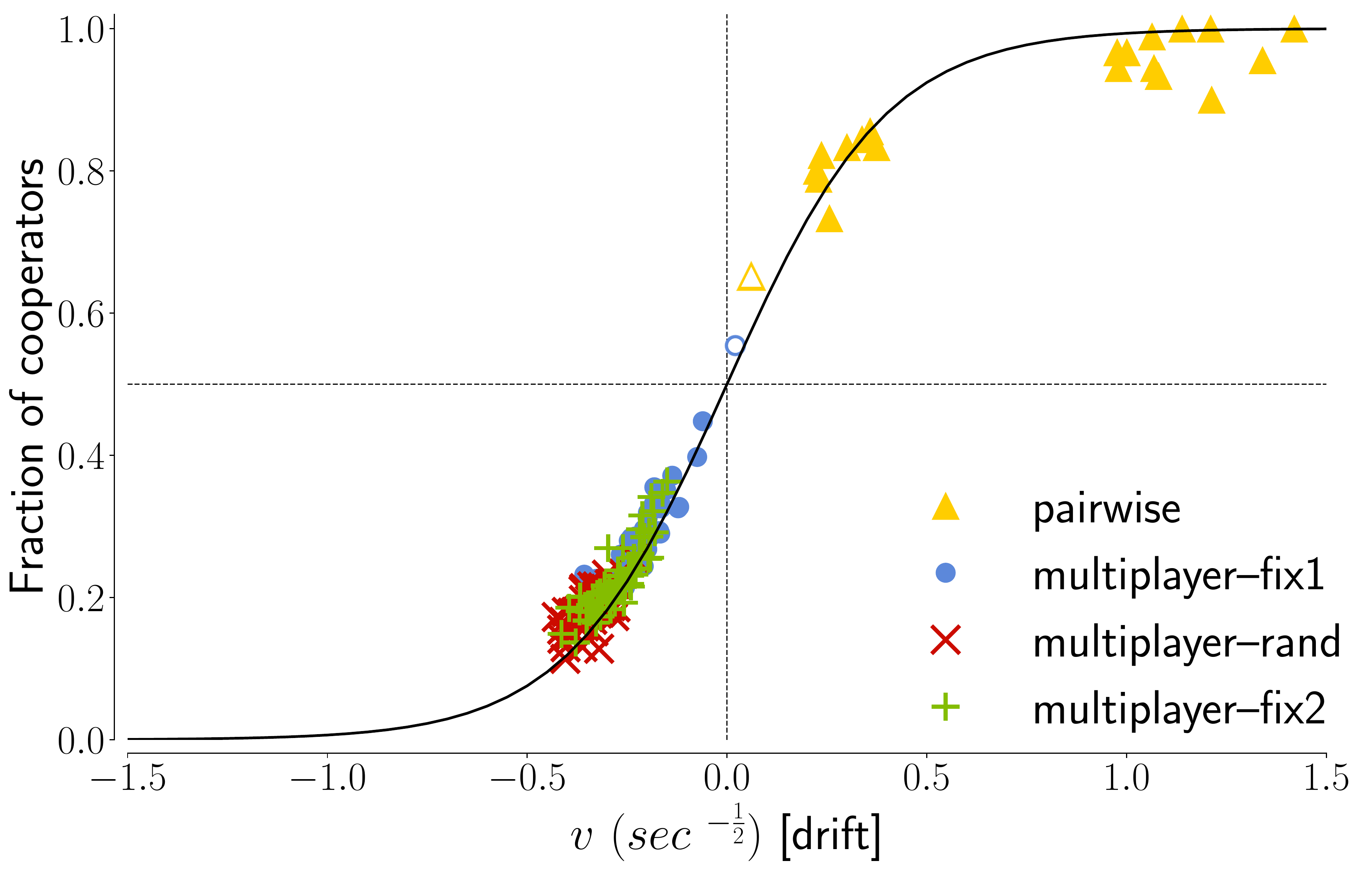}\\
\raisebox{2.5cm}{(c)} \includegraphics[angle=0, width=0.45\textwidth]{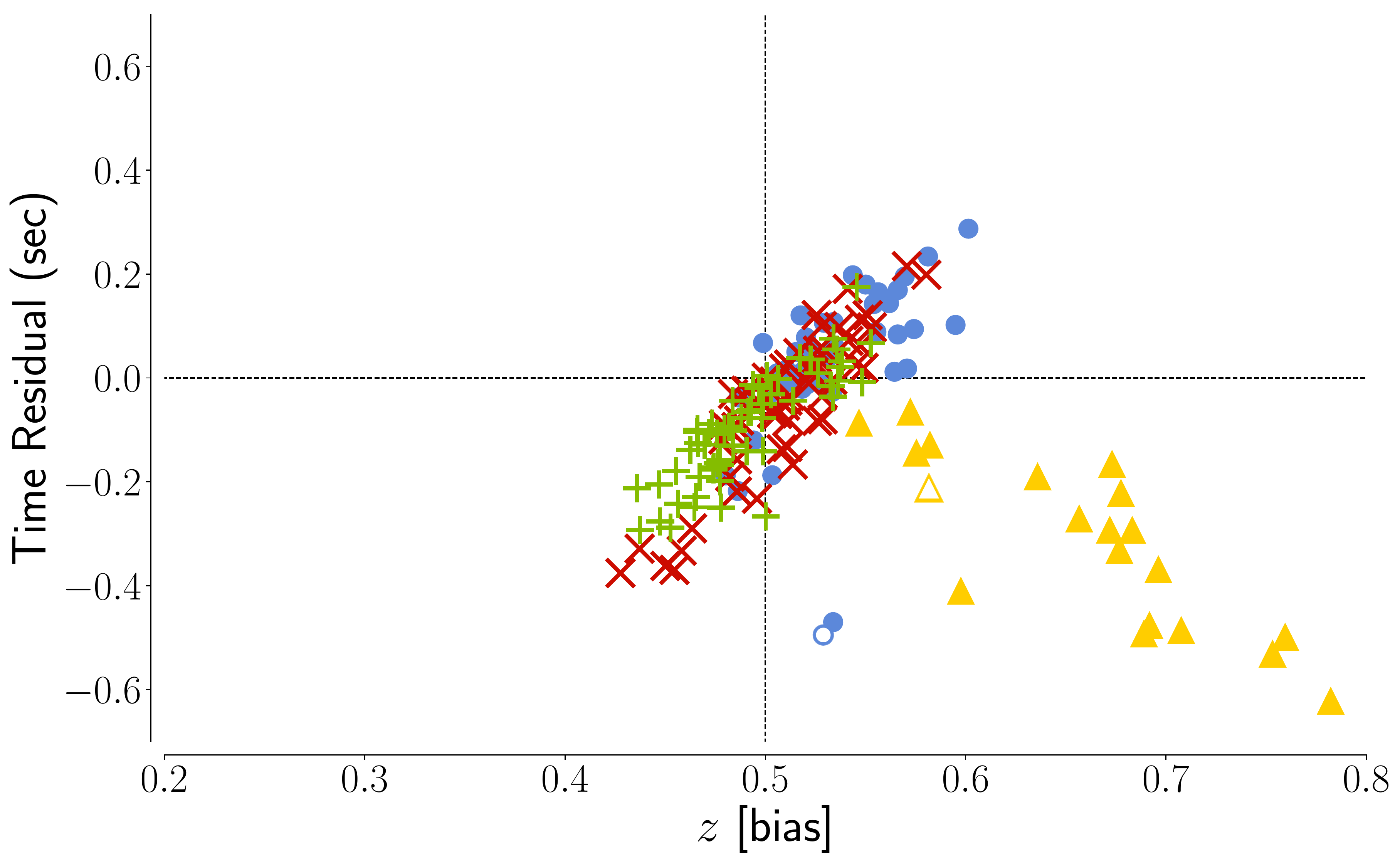}&
\raisebox{2.5cm}{(d)} \includegraphics[angle=0, width=0.45\textwidth]{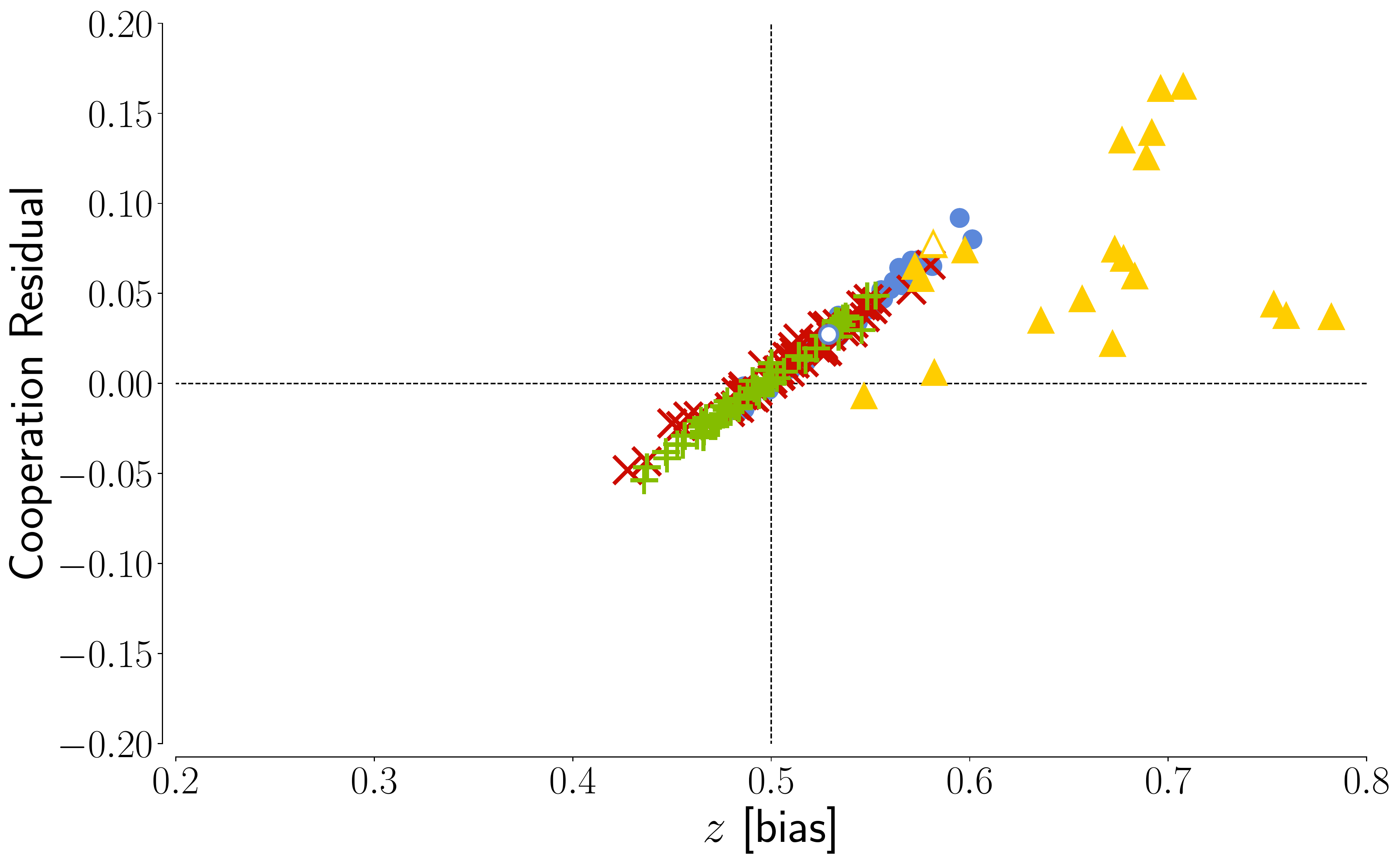}\\
\end{tabular}
\caption{
{\bf How the DDM parameters are linked to average decision time and cooperation level in game theory experiments.}
{\bf (a)} The dominant parameter for the decision time is the threshold $a$. The black solid line represents the value $\langle t-t_0 \rangle=a^2/4$ expected in absence drift and bias ($v = 0$, $z = 0.5$).
{\bf (b)} The dominant parameter for the cooperation level is the drift speed $v$. The black solid line represents the value $C^{teo}(v,a,z=0.5)=(1+\exp(-a v))^{-1}$ expected in absence of drift ($z=0.5$), where we fixed the threshold at the initial value $a = 5$ sec$^{\frac{1}{2}}$.
{\bf (c)} The drift $z$ plays a secondary role in the decision time: its values is positively correlated with the residual $R_t = \langle t \rangle - \langle t \rangle^{teo}_{z=0.5}$ if $v < 0$ (as in the multiplayer experiment) and negatively if $v>0$ (as in the pairwise experiment). For the multiplayer experiment, where we have better statistics, the value of the residual is an order of magnitude less than the average decision time.
{\bf (d)} The drift $z$ plays a secondary role also for the cooperation level: its values is positively correlated with the residual $R_C =  C - C^{teo}(v,a,z=0.5)$. In the multiplayer experiment, the bias alters up to the $10\%$ of all decisions, an effect of the same order of magnitude of the finally attained cooperation level of $\approx 20\%$. 
}
\label{figure4}
\end{figure*}

The value of $a$ is the main factor determining the average decision time. In figure~\ref{figure4} (a) we show how our experimental results align along the theoretical relationship $\langle t \rangle = \frac{a^2}{4}$  valid in absence of bias and drift ($z = 0.5$, $v=0$, see Methods). The value of the average decision time in our experiment is thus dominated by the caution with which the players consider their response. The decisions become thus progressively quicker mostly because less information is demanded for the final deliberation. We emphasise that the fit is done simultaneously for both decisions, therefore the observed evolution of the threshold parameter equally influences RT for both decisions.


{\bf Drift} -- corresponds to deliberation. In the first round of both scenarios the drift speed $v$ is zero (Fig.~\ref{figure3} (b)), which is consistent with the lack of information to consider for a rational deliberation. It then progressively diverges to positive values (towards cooperation) for the pairwise interaction and to negative values (toward defection) for the multiplayer game. These opposite trends directly reflect the level of cooperation finally attained at the end of the two experiments: the game experience is therefore providing evidence suggesting cooperating among cooperators and defecting among defectors. 

With the notable exception of the first round, we can note in Fig.~\ref{figure4} (b) that the variability of $v$ is here the main factor behind the evolution of the cooperation rate $C_R$, with  $C_R\approx  (1+\exp{\left(-5v\right)})^{-1}$, as expected for no bias ($z=0.5$) and with the threshold observed at the first round  ($a=5$ sec$^{\frac{1}{2}}$). In our iterated game, cooperation or defection therefore emerge here mostly as a consequence of informed deliberation.

The different scenarios explored in the multiplayer experiment allow us to also to illustrate how the absolute value of $v$ depends on the difficulty of the task~\cite{Smith:2004jo}. Indeed, $|v|$ progressively grows between fix1 and fix2, while for the random phase the value is stable and, on average, significantly higher than in fix1 and fix2 (see Supplementary Table 1). These higher values of the drift speed $|v|$ are consistent with the easier design of the random phase, where the information on the previous decisions of the neighbours is not available. At the same time, deliberations in the similarly designed fix1 and fix2 become gradually easier, as the players become more and more proficient in processing the information provided.
In the random phase of the multiplayer experiment the absolute value of the parameter $v$ is higher because the neighbours are changing  and therefore they do not need to keep track of all relationships.

\begin{figure*}[ht!]
\centerline{
\includegraphics[angle=0,width=0.9\textwidth]{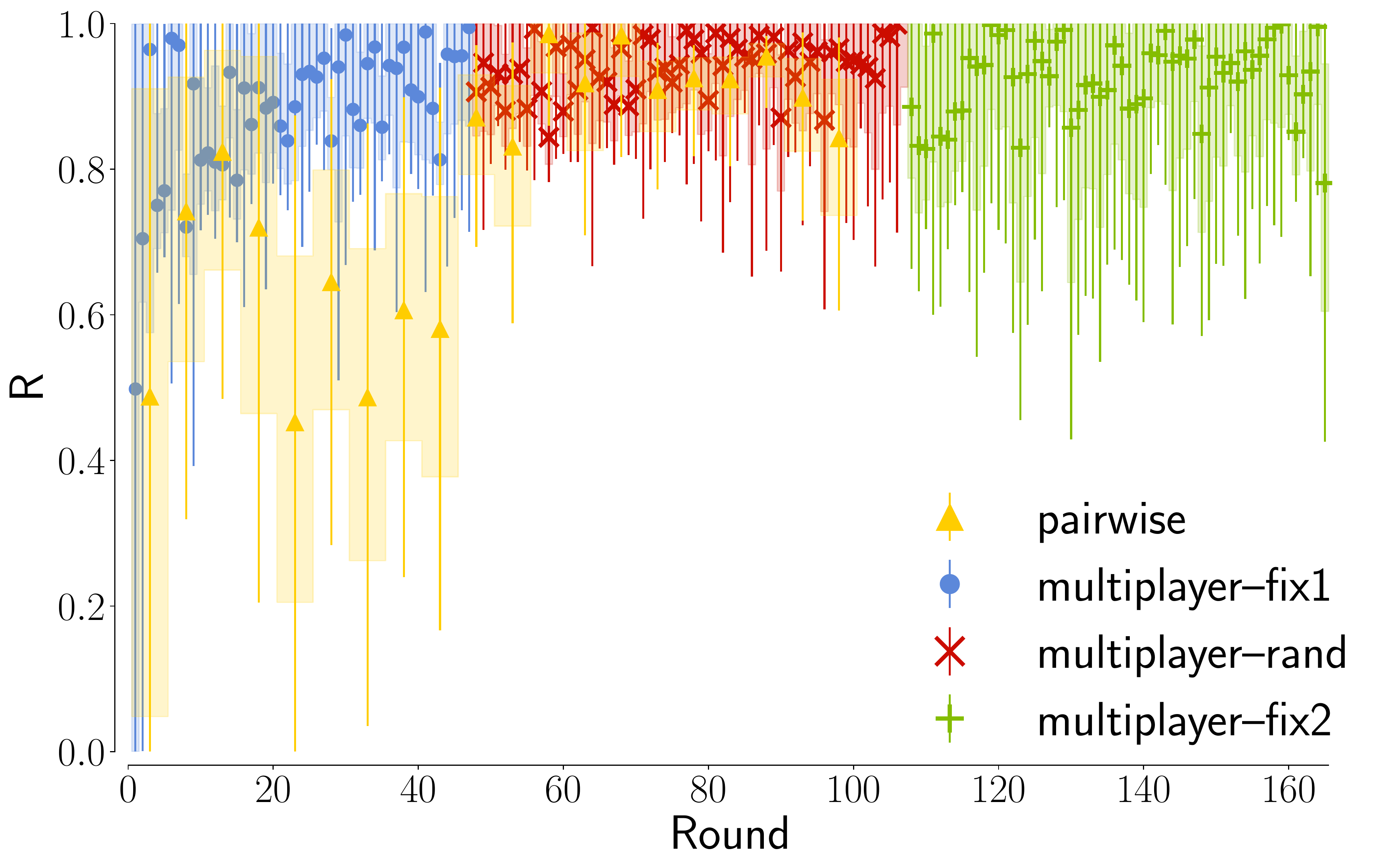}
}
\caption{
{\bf The players' choices becomes progressively dominated by deliberation as the experiments progress.} In figure we represent the `rationality ratio', a quantity derived by the cooperation levels designed to be $R=0$ for totally intuitive decisions and $R=1$ for totally rational decisions.
Tracking the evolution of $R$ we can see how, after a sufficient number of repetitions, the decisions made become mostly a consequence of rational deliberation. For the first rounds our fits suggest an almost perfect balance between intuition and deliberation ($R\approx 0.5$). At round 20 in the multiplayer experiment, and at round 50 in the pairwise experiment, the ratio $R$ settles to values higher than 0.8, indicating a stable rational behavior. Error bars represent here the variability due to the uncertainty in our estimate of $v$ (see Fig.~\ref{figure3} (b)).
}
\label{figure5}
\end{figure*}

{\bf Bias} -- tells us about the initial inclination people might have towards one of the two options. The initial value of the a-priori bias $z$ is $\approx 60\%$. This value perfectly reflects the cooperation rate in the first turn: we have indeed $C_R = z$ when the drift is null ($v=0$), as we have in the first turn. In Fig.~\ref{figure3} (c) we observe how the bias then grows in the pairwise experiment (where players cooperate) and drops in the multiplayer experiment (where they mostly defect). This drop is independent from the players' strategy (see Supplementary Fig. 5). This suggests that any inclination prior to rational deliberation is strongly influenced by the other players' behavior, with the bias increased following positive interactions as much as reduced after negative ones. The initial bias towards cooperation is however resilient, as in the multiplayer experiment it resets to $\approx 60\%$ after each pause. 
The monotonic decreasing trend is significant in each of the three phases in the multilayer experiment, if we exclude the last ten rounds of fix2, as the Mann-Kendall Test for a decreasing trend is passed with probability $0.04$, $1.5 \cdot 10^{-5}$ and $3 \cdot 10^{-4}$ respectively. The trend is instead not perfectly monotonic for the pairwise experiment, and if we consider the last part of fix2.
The sawtooth shape of the evolution of $z$ is not reflected in the cooperation rate, because the drift quickly becomes, and remains, the dominant factor in the decision process (see `Rationality' and Fig.~\ref{figure5}). The bias parameter mostly captures minor deviations of decision time and cooperation rate from the analytical tendencies described above (Fig.~\ref{figure4} (c) and (d)).


{\bf Non-decision time} -- corresponds to the time before the decision process starts to happen (for example, before the participants realize that the new round started or the time they use to press the button on the mouse etc). In Fig.~\ref{figure3} (d) we show that the perceptual and motor processes associated with the task speed up with repetition, but a stable value is reached after about 5 rounds. In the simpler pairwise experiment the non-decision time reaches the value expected for a purely visual reaction time of $t_0\approx 0.3$ seconds~\cite{shelton2010comparison}. The initial drop in $t_0$ partially accounts for the decrease in RTs over time (Fig~\ref{figure1} (left)). Notably, the drop becomes significant after the pauses in the multiplayer experiment. Excluding the first 5 rounds, the average non-decision time is of $0.29\pm0.03$ in the pairwise experiment and $0.59\pm0.03$ in the multilayer experiment, for which also the first rounds of rand and fix2 has been filtered out. In principle, we could have assumed non-decision time as a constant value across all rounds of a given experiment, but the large difference of the first rounds' values we observed here suggests that it is here relevant to keep this fourth parameter in the fits.


{\bf Rationality.} 
In Figure~\ref{figure5},  we represent the evolution of the `rationality ratio' $R$ we introduced in Eq.\ref{eq_R}.
In case of totally intuitive decisions one would expect $R=0$ while in case of totally rational decision $R=1$. In our experiments, the numerical result for the first rounds in both experiments suggests a compromise between rational and intuitive behavior ($R\approx0.5$). It is however important to remark here that for both experiments the value $v=0$ lies within the margin of errors for the first rounds, and consequently our results are not inconsistent with the hypothesis that the first round are dominated by intuition ($R=0$). If for the first rounds we have a balance between intuition and deliberation, it then appears clear that as the players learn how to play the game, their behavior becomes ultimately rational as $R$ grows reaching values close to 1.

These results show how rational deliberation, based on the gathered information, quickly becomes dominant over an initial intuitive bias towards cooperation. The role of intuition is still decisive when any information on the expected behavior of other players is lacking, as in the first rounds or in one-shot experiments~\cite{cappelletti2011being,rand2012spontaneous,rand2014social} where the faster decision is the cooperative one. Consistently, in interative experiments, which are averaging the first round with a relatively large number of the later rounds~\cite{piovesan2009fast} this trend has not been observed.

Intuition is not however strictly suggesting cooperating, as the bias can turn towards defection after playing multiple rounds in an un-cooperative environment. Nevertheless, the resilience of the bias we observe between the phases of the multiplayer experiment confirms that we might have a `natural' intuitive optimism concerning the social behavior of the others. This is because, although the initial attitude leans toward faster decisions associated with cooperation, learning the game implies becoming a rational player.  If the rational choice is to defect, as in our multiplayer experiment, spontaneous altruism is just a transient effect possibly due to optimistic initial expectations~\cite{fehr2003nature}, after which it defection becomes the quicker answer.

\section*{Discussion}

Our results above are not only proof that the use of the Drift Diffusion Model can be extended to the description of the complex strategic actions taken during game theory experiments, but illustrate how an accurate modelling of decision times allows us to get new detailed insight on human decision process from a neuro-economics~\cite{Glimcher:2004et} perspective.
This kind of statistical approach is possible when the datasets are large enough to make it possible to statistically test the exact shape of the empirical decision times distribution. In our case, the multiplayer game with 169 players allowed us to study every round separately to confirm the remarkable ability of such a simple model to summarise very rich experimental results with up to 8000-10000 decisions. Tracking the evolution of the drift speed $v$, the accumulation threshold $a$, and the initial bias $z$ during the experiment offers a new perspective into this learning process. This would not be possible by only studying the average decision times and permits a new interpretation of differences between experiments, contexts, experience, inclinations and strategies.

In particular, we show here that analyzing the empirical results using DDM is a method surely more appropriate for the difficult task of distinguishing between deliberation (described by the drift) and intuition (associated to the bias) than simply comparing the average response times. We see here that it is not in general true that faster decisions are dictated by intuition, nor that the intuition necessarily suggests cooperating.
The drift in the DDM embodies the effects of strength-of-preference~\cite{krajbich2015rethinking}, as it can be seen as representing how the different utility between the chosen and unchosen options is estimated during deliberation~\cite{Busemeyer:1993bm}.
When the context of the game suggests the players to defect, cooperative decisions are faster only in the first round, or equivalently in one shot games. In this case, the decision is indeed dictated by intuition, since there is no game experience to base a rational decision on, and the cooperation level equals the value of the bias $z$.
But if then the decision process is repeated, rational deliberation becomes dominant over an initial intuitive bias towards cooperation. As we learn the game, the value of $|v|$ grows as the process of integration of evidence becomes more efficient, allowing us to make faster but more rational decisions. This process is coupled with a progressive drop in the threshold $a$, as players also become less cautious and demand less information for the final deliberation, and with a rapid accommodation of our perceptual and motor systems to the task at hand embodied by the non-decision time $t_0$. 
The transition between naive and informed decisions demands only a few rounds (See Fig.~\ref{figure5}), after which the effect of the initial bias towards cooperation becomes marginal.
At this point, the players decision are no longer naive but mostly rational, however the underlying bias also is subject to an evolution which depends on the context of the game.
When we play a game where most of the players are defectors, the intuitive decision progressively becomes to defect, while when we play with cooperators we become even more biased towards cooperation. The good news is that after only a short pause, the bias towards cooperation resets to its initial `natural'  value of 60\%. This `natural' bias is therefore resilient to short term experiences. These phenomena we observed here for Prisoner's Dilemma can be tested with different experimental setups, such as different games and possibly include tasks of different complexity and time-pressure, with the only design constrain being a large number of repetitions.

More in general, the possibility of monitoring in real time the learning process by studying a single and easy to measure quantity, response times, allows for the development of new practices to test whether the players (humans or animals) have mastered the task and to eliminate any initial transients by verifying if they have reached a steady state.
Future studies can take advantage of the apparent universal validity of DDM, the range of use of which spans from memory retrieving, to perceptual, value based, and strategic decision making. Indeed, this suggests that the neurophysiology behind all these types of decision is similar, and that a set of decision models with a more elaborate architecture, which can be re-conducted to DDM under some parametric choices, represent natural candidates for a more detailed description of strategic decision making~\cite{Bogacz:2006fj}.

\section*{Materials and Methods}

\subsection*{Multiplayer experiment}
The game theory experiment was performed in April 2009, at the Universidad Carlos III de Madrid in Spain.
The participants were 169 students from the Engineering Campus of Leganes. The subject's age ranges between 18 and 26 years old. They played a Multiplayer Iterated Prisoner's Dilemma game on a square lattice, simultaneously with eight neighbours (Moore's neighbourhood: up, down, left, right and diagonally). The lattice had periodic boundary conditions as if the players were placed on a torus. They had to choose one action: Cooperate (C) or Defect (D), which would be then applied to the game with each of their eight neighbours. For each neighbour, they would earn the payoff which is presented in the table:
\begin{table}[h]
\begin{center}
\ra{1.3}
\begin{tabular}{c|cc}
    & \bf{C} & \bf{D} \\
    \hline
    \bf{C} & 7 & 0 \\
    \bf{D} & 10 & 0 \\ 
\end{tabular}
\end{center}
\end{table}

If they both cooperate, they each earn a ``Reward for mutual cooperation'' $R=7$ cents, if they defect and the other person cooperates they  earn a ``Temptation to defect'' $T=10$ cents, and if the other player defects they earn nothing. Therefore ``Punishment for mutual defection'' and the ``Suckers payoff'' are both zero $P=S=0$. Notice that this is a so-called weak Prisoner's Dilemma, meaning that ``Punishment'' and ``Sucker's payoff'' are here equal. If the other player defects, the focal player earns nothing in any case and there is no cost in switching to cooperating. The reason for choosing a weak Prisoner's Dilemma setup is to make the game more cooperative than a strict Prisoner's Dilemma. In the original experiment, which was designed to test if the cooperation could be established, it was important to set the conditions as favorable to cooperation as possible. Here, this particular choice is however not relevant. For each round, the total payoff for each player is then calculated as a sum of the outcomes from all eight games. At the end of the experiment, the participants are rewarded proportionally to the payoff accumulated in Euros.

All participants experience the following steps: i) login; ii) read the instructions; iii) test if they understood the instructions; iv) the experiment itself. The experiment has three phases: fix1 (fixed network during all the rounds), rand (network reshuffled after every round), fix2 (fixed network, however different from the network in the fix1).
In the original paper, these phases were called exp1, control and exp2. Here, we are renaming them in order to provide more intuitive names for this specific context. 
The size of network was a very important factor for the original experiment. In particular, the objective was to have a number of players an order of magnitude larger than any previous experiment. Because of the number of participants needed, it was impossible to have multiple sessions. However, we believe that the system is large enough to self-average, as a single realisation of a large system can be equivalent to averaging over a whole ensemble.
In any case, the lack of multiple session does not have any influence on the particular analysis in this paper.
At all stages, the information provided to them on the screen (Supplementary Figures 1 and 2) is the result of the previous round they played: the actions and payoffs of themselves and all of their neighbours. This means that, in the rand phase, the players have no information about their new neighbours when they need to make the decision and they only rely on their experience and expectations.
There was no practice round of the experiment, as it was not intended for the participant to receive any experience in the game before it started. However, we did make sure that they understood the rules by giving them a 4 different situations which could occur during the game and asking them how much money they would earn in those situations. This was done to ensure that each participant understood the game well, given that it has been shown that almost 30\% of participants of participant will not read the instructions carefully \cite{oppenheimer2009instructional}. Understandably the particular choice of the examples could frame the participants, however the risk of them not understanding the instructions was far greater and the examples were chosen to be balanced and the least likely to frame the participant in one way or the other. It should be also emphasised that all students participated in all three phases of the experiment. Therefore the different phases are not independent, since in each phase they have the experience from the previous one. Consequently, the phase we call rand is not a proper control of the experiment, in which the player without the previous experience would play the game. In order to avoid having players defect in the last round, knowing that there is no more game to play and therefore no reason to cooperate anymore, the end of the game was decided randomly by computer and it was at round 47, 60 and 58 respectively for fix1, rand and fix2. The data set is very large, making in total more than 27000 decisions. The software used in the experiment was developed in PHP, javascript with python controlling the background processes. The participants had 30 seconds to take an action afterwards the computer would make a decision for them, but there is no countdown to the deadline shown.
The automatic decisions were 
excluded from the analysis.  As the primary purpose of the experiment was not a precise measurement of the decision times, they were recorded on the server, not on the client. Therefore there was a certain delay in their measurement.  We estimate these delays by analyzing the times recorded for the automatic decisions, which are made by the client in exactly 30 seconds. This recording goes under the same procedure as the other times. 
On average the delay in the system was less than 0.03 seconds, which makes it only $\approx$0.5\% of the average decision time. 
More importantly, the standard deviation of the delays is even smaller of around only 0.01\% ($4\cdot 10^{-4}$ seconds).
Thus, their influence on the shape of the decision times distribution is only of a negligible shift that gets included in the non-decision time $t_0$.
Three outliers, with delays of more than 30 seconds, were excluded from the analysis.

\subsection*{Pairwise experiment}
This experiment has been performed in 2015 at Brussels Experimental Economics Lab, at Vrije Universiteit Brussels. The requirement site is made using ORSEE \cite{greiner2004online}. The game played was Weak Pairwise Prisoner's Dilemma; each player played with one fixed partner a Prisoner's Dilemma with the following payoff matrix:  
\begin{table}[h]
\begin{center}
\ra{1.3}
\begin{tabular}{c|cc}
    & \bf{C} & \bf{D} \\
    \hline
    \bf{C} & 3 & 0 \\
    \bf{D} & 4 & 0 \\ 
\end{tabular}
\end{center}
\end{table}

The participants were mostly students aged between 19 and 32, with the majority being students. There were 18 players playing in 9 independent couples for 100 rounds each. They each had 30 seconds to take the action and they had a counter on the screen telling them how much time they had left. However after 30 seconds were up, they could still play, but would be given a warning to take action immediately. In most rounds all the players took their actions in a much shorter time than the given 30 seconds limit. The experiment lasted less than 45 minutes from the moment participants entered the room until the moment they left. We adapted the software from the previous experiment to be used for this one. 

\subsection*{Ethical Statement}

Both experiments were performed according to the standards of socio-economics experiments. The participants' anonymity was always preserved and they were never deceived during the experiments. All participants signed an informed consent. For the multiplayer experiment, please take a look at the Ethics statement in the original paper~\cite{grujic2010social}. For the pairwise experiment, ethical approval by reference number ECHW2015\_3  was obtained from the Ethical Commission for Human Sciences at the Vrije Universiteit Brussel to perform this experiment.


\subsection*{Drift Diffusion Model}

For the description of the response times distributions for binary decisions, we use a statistical decision model, the Drift Diffusion Model (DDM), which has the advantage of  being defined by a simple linear, first-order, stochastic differential equation. In the DDM, at each moment subjects randomly collect evidence in favour of one of two alternative hypotheses. The continuous integration of evidence in time is described by the evolution of $x(t)$ as a one-dimensional brownian motion with diffusion coefficient $\sqrt{D}$ and a drift $v$:
\begin{equation}
dx = v dt + \sqrt{D} \xi(t)
\label{eq:1DbrownDrift}
\end{equation}
For each $dt$ the quantity $x(t)$ is increased by $v dt$ (drift term) plus a noise $\sqrt{D}\xi(t)$ (diffusive term), where $v$ and $\sqrt{D}>0$ are constant and $\xi(t)$ is a white noise. In absence of boundary effects, the probability density $P(x,t)$ of the solutions of Eq.~(\ref{eq:1DbrownDrift}) is normally distributed with mean $\mu = x(0) + v t$ and variance $\sigma^2 = Dt$.

Given two barriers at $x=0$ and $x=a$ and an initial condition $x(0) = z\cdot a$ with $z \in (0,1)$, it is well defined the commonly called ``gambler's ruin problem''~\cite{feller1968introduction}, where $x(0)$ 
represents the initial bankroll of the gambler, the absorption at $x=a$ represents the gambler leaving a possibly unfair game (if $v\neq 0$)  after collecting her target winnings $a$, and the absorption at $x=0$ represents the gambler's ruin. The probability distribution of the times at which the process reaches the origin $x=0$ before reaching the exit value $x=a$ is known as F\"urth formula for first passages
 \begin{align*}
P(t;v,a,z,D) &= \frac{\pi \sqrt{D}}{a^2} \exp\left(-  \frac{v z a}{\sqrt{D}} - \frac{v^2 t}{2\sqrt{D}}  \right) \\ &\times \sum_{k=1}^\infty k \exp\left( -\frac{k^2\pi^2\sqrt{D}t}{2a^2}\right)\sin\left({k\pi z}\right) \, ,
\label{eq:DDM_bias}
\end{align*}
which represents the probability distribution that a gambler will be ruined at time $t$ and is characterised by an exponential tail. 

The parameters $v$, $D$ and $z$ being interdependent, it is common practice to set $D=1$ and use only the other three parameters to fit to the data with the curve $P(t;v,a,z) = P(t;v,a,z,D=1)$. This simplification is equivalent to re-defining the process in Eq.(\ref{eq:1DbrownDrift}) as $dx' = \frac{v}{\sqrt{D}} dt +  \xi(t) $, for a rescaled quantity $x' =x/\sqrt{D}$, which starts at $x'(0) = x_0/\sqrt{D}$ and is interrupted when $x' = a/\sqrt{D}$ or $x' = 0$.
By imposing $D=1$ we have, for dimensional reasons, $[a] =  [t^\frac{1}{2}]$ and $[v] = [t^{-\frac{1}{2}}]$, which is consistent with the dimensions of the rescaled constants $v'$ and $a'$ (note that the dimension of D is sec$^{-1}$).

For describing decisions in a Prisoner's Dilemma experiment, we associate here the barrier at $x = 0$ to defection and the barrier $x = a$ to cooperation. When $v$ is positive, the gathered information for the deliberation is tendentially in favour of cooperation, while when it is negative the evidence gathered is mostly supporting defection. The module of the drift speed $|v|$ [sec$^{-1}$] is the signal-to-noise ratio of the drift representing the unbalance in the amount of evidence supporting the two alternative options. $|v|$ depends upon the difficulty of the decision. The lower $|v|$ the more difficult the task~\cite{Glimcher:2003cy,Smith:2004jo,Churchland:2008ds}.
The initial condition $x(0)$ here describes the biases prior to deliberation.  An unbiased initial condition would be $z = 0.5$. For values above $0.5$ the decision maker has a bias toward cooperation and below $0.5$ the initial bias is toward defection. The probability distribution $P(t;v,a,z)$ thus describes the decision times of rounds where the ultimate decision is defection. Conversely, the probability distribution to associate with cooperative rounds is $P(t,-v,a,1-z)$. The area under the curve $C_R^{teo}(v,a,z) = \int_0^\infty P(t;-v,a,1-z) dt$ corresponds to the fraction of cooperation expected. The curve fit algorithm we used~\cite{wiecki2013hddm} fits at the same time the non-normalised distributions $P(t,v,a,z)$ and $P(t,-v,a,1-z)$, thus evaluating at the same time the decision times for cooperation and defection, and the fraction of cooperation.

In the modelling of decision times, the quantity $x(t)$ is associated to the difference in the amount of evidence supporting the two alternatives. To better illustrate this point of view, it is convenient to shift the process so that an unbiased initial condition, which lies exactly between the two barriers $x(0) = a/2$ (z=0.5), is mapped to $x=0$ and the barriers become symmetrical $x = \pm a/2$~\cite{Bogacz:2006fj}. The new quantity $x' = x- a/2$ can then be associated to 
i) the difference in activity (firing rate) in two neuronal populations, each associated to one of the two alternative options (neuroscience perspective~\cite{Smith:2004jo});
ii) the difference between the utility expected for the two options (economic perspective~\cite{Busemeyer:1993bm}). 
In both cases, $|x'|$ represents the expected benefit obtained from taking the decision which is better supported by the evidence. Again from the neuroscience perspective, the final decision is then made by a second brain circuit that acts downstream detecting the event of the threshold crossing~\cite{Lo:2006bq}.

The distance between the barriers $a$ is the threshold parameter of the DDM, representing the amount of information needed for the final decision. Its value can be naturally associated to the average first passage time of the diffusion process without drift ($v=0$) in the unbiased case ($z = 0.5$). 

Indeed, the average value of the decision times $\langle t\rangle$ for this unbiased case has the form~\cite{Bogacz:2006fj}
\begin{equation}
\langle t \rangle^{teo}_{z=0.5} = \frac{a}{2v}\tanh{\left( \frac{av}{2}\right)} \,.
\label{averageT}
\end{equation}
In the limit for $v\to0$, we have a pure diffusion process and Eq.(\ref{averageT}) leads to the identity $\langle t \rangle = a^2/4$ (see Fig.~\ref{figure4} a). Therefore, we can identify $a^2/4$ as the average first passage time for a similar diffusion process without drift. 

Again for the unbiased case, it is possible to compute the fraction of cooperation expected given $v$ and $a$ ~\cite{Bogacz:2006fj} (see Fig.~\ref{figure4} b):
\begin{equation}
C_R^{teo}(v,a,z=0.5) = \frac{1}{1+\exp{\left(-av\right)}} \, . 
\end{equation}

Higher values of $v$ yield faster and more precise decisions, whereas higher values of $a$ permit averaging out uncorrelated noise more consistently, producing slower but more accurate responses. As a consequence, the model is able to describe consistently the speed/accuracy trade-off  observed for decisions made under pressure~\cite{Milosavljevic:2010wi}.

\subsection*{Model fitting}

The data were fit using an open-source python library, the HDDM~\cite{wiecki2013hddm}, based on a hierarchical Bayesian estimation of the four free DDM parameters, which are all and the only parameters included in our analysis.
The decision times and the associated action were indeed fit by pooling data across participants, at round level in the multiplayer experiment (169 response times), or aggregated over five consecutive rounds in the pairwise experiment (90 response times). We are therefore fitting the model to the group as if the are all copies of an `average subject'. This assumption is of course limiting, but our choice is motivated by the need of proposing a method that can be scaled to what is commonly done in Game Theory Experiments, where the task is repeated only few times, and often only once.

\section*{Acknowledgements}
We thank Marc Barthelemy, Paul Brodersen, Jose A. Cuesta, Tom Lenaert, Valentina Lorusso, and Angel Sanchez for comments on an early draft; Federico Corradi and Pierluigi Sacco for useful discussions; Letty Ingrey for proof reading the manuscript.

\section*{Author contributions}
RG modelled the data and prepared the figures. JG was part of the team who designed and performed the experiment and preprocessed the raw data. Both authors discussed the results and its implications at all stages and wrote the manuscript together.

\section*{Competing interests}
The authors declare no competing interests.

\section*{Data availability statement}
Anonymised data are available from the authors upon request.

\clearpage

\bibliographystyle{naturemag-doi}
\bibliography{decisionTimes.bib}

\clearpage

\section*{Supplementary Figures and Tables}

\setcounter{figure}{0} 
\setcounter{table}{0} 


\begin{figure*}[ht!]
\def\figurename{Supplementary Figure}
\centering
\includegraphics[width=0.48\textwidth]{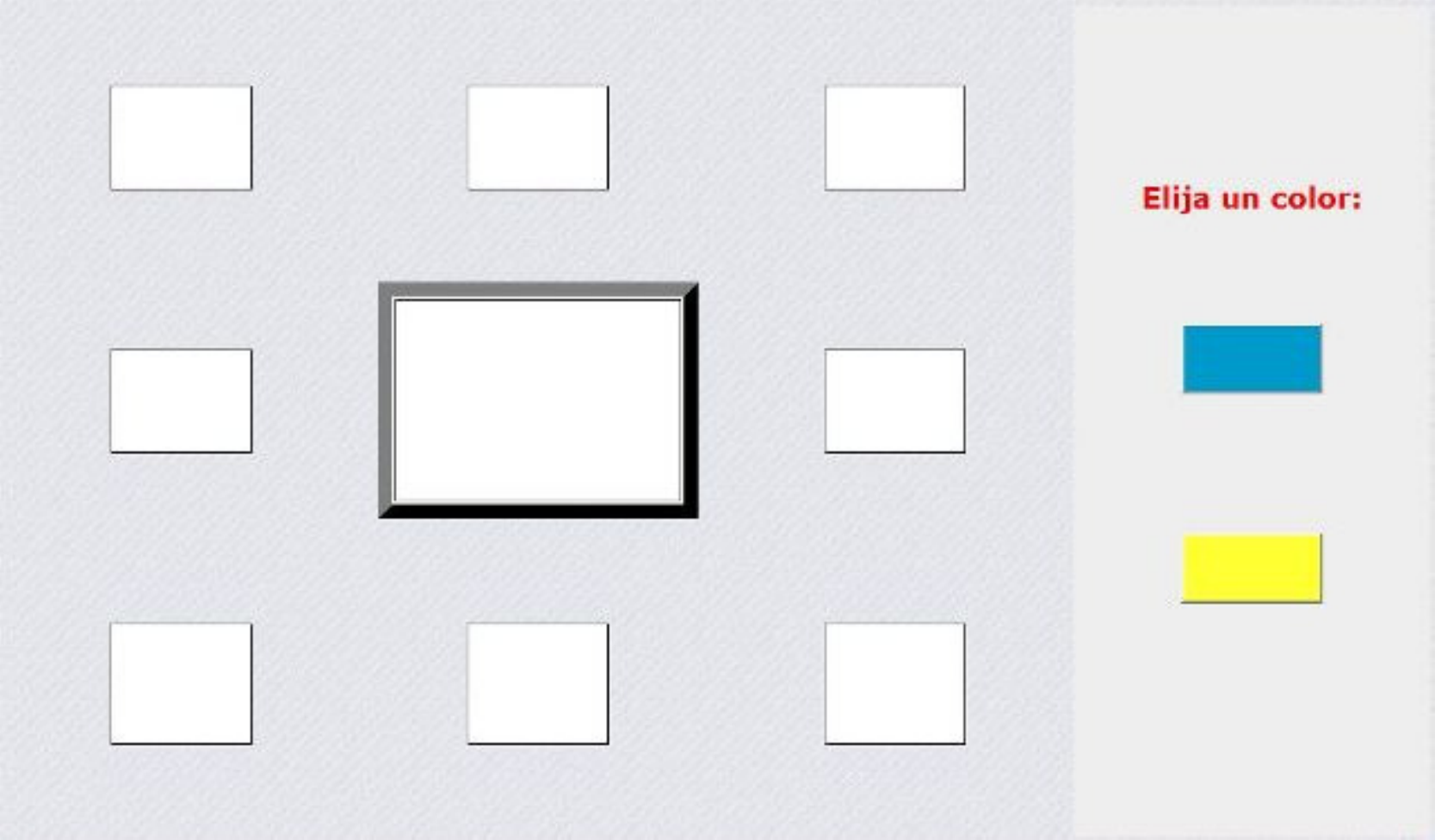}
\qquad
\includegraphics[width=0.455\textwidth]{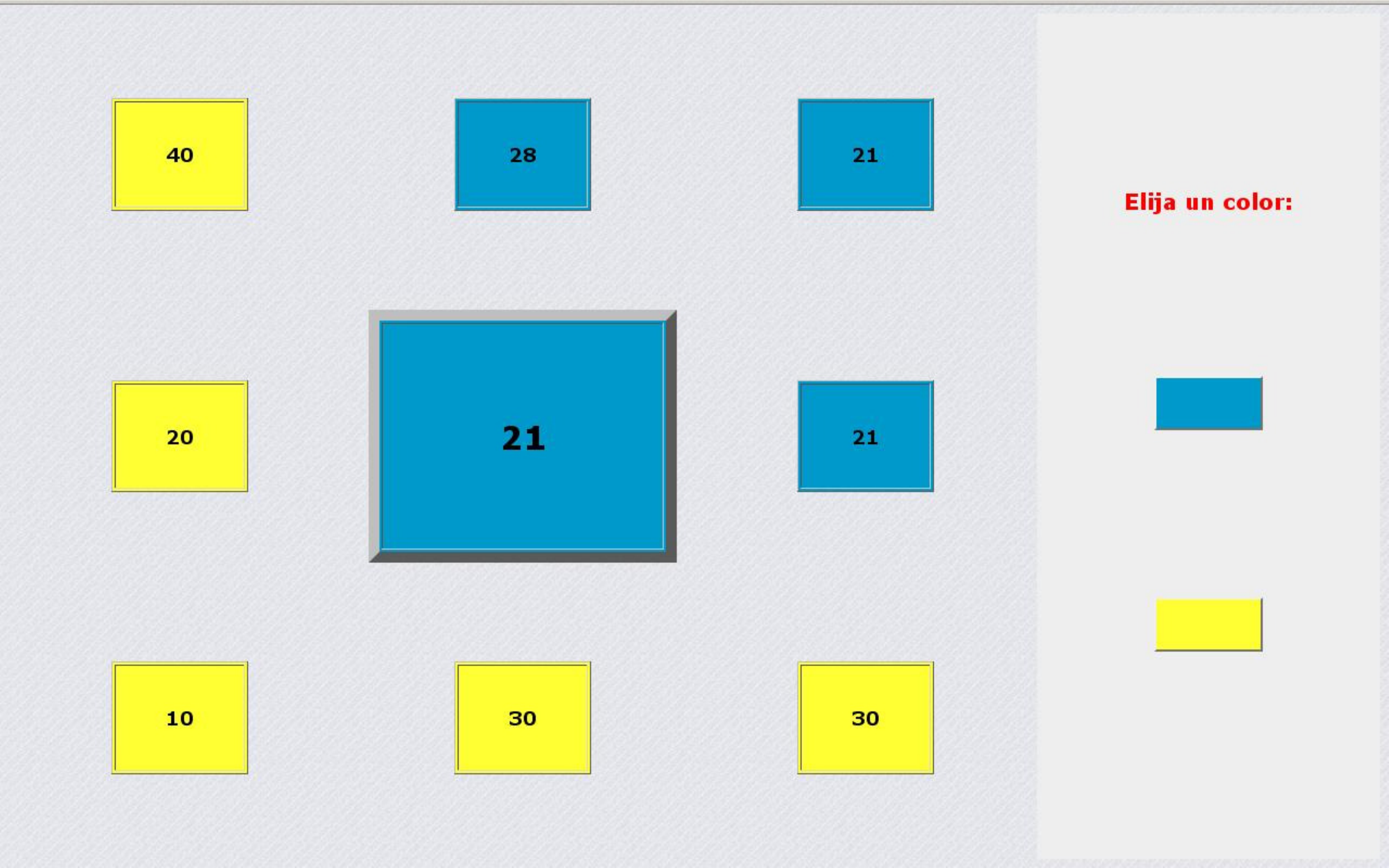}
\caption{{\bf The interface of the multiplayer experiment.} On the left: the screen in the first round of each phase. On the right: the screen in the following rounds.}
\label{SI:figure1}
\end{figure*}

\begin{figure*}[ht!]
\def\figurename{Supplementary Figure}
\centering
\includegraphics[width=0.47\textwidth]{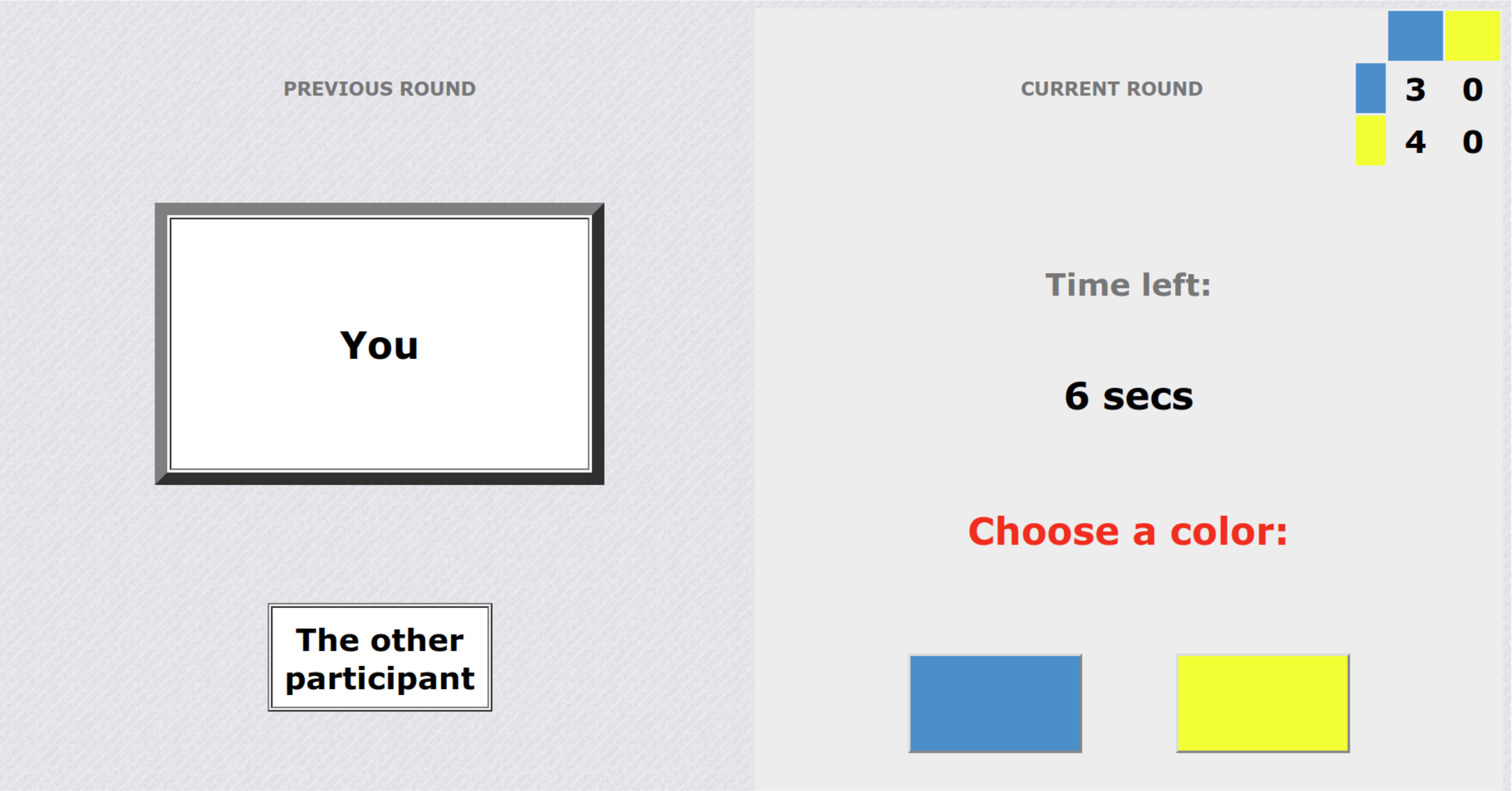}
\qquad
\includegraphics[width=0.47\textwidth]{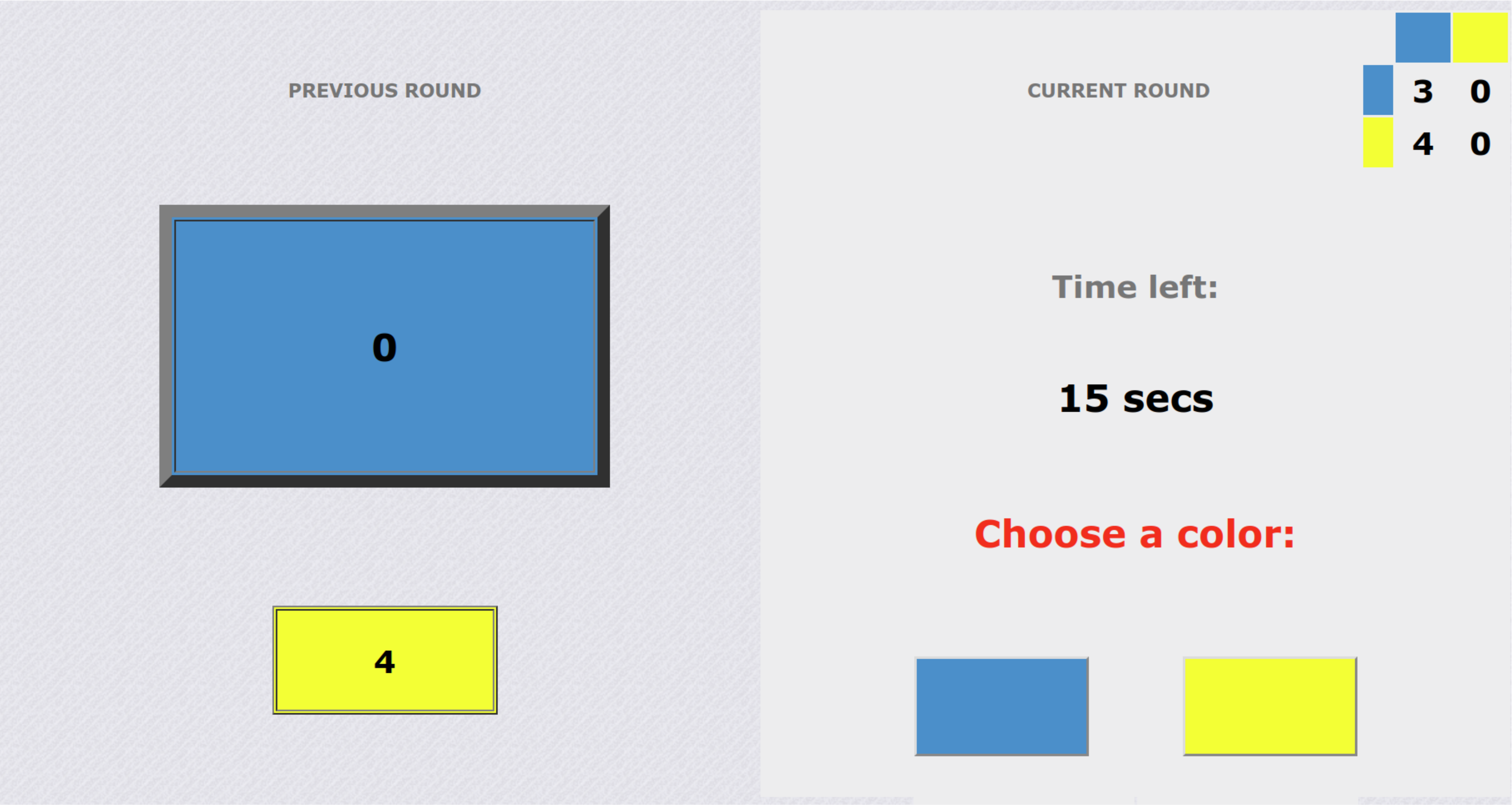}
\caption{{\bf The interface of the pairwise experiment.} On the left: the screen in the first round of each phase. On the right: the screen in the following rounds.}
\label{SI:figure2}
\end{figure*}

\begin{figure*}[ht!]
\def\figurename{Supplementary Figure}
\begin{tabular}{cc}
\raisebox{2.5cm}{(a)} \includegraphics[angle=0, width=0.45\textwidth]{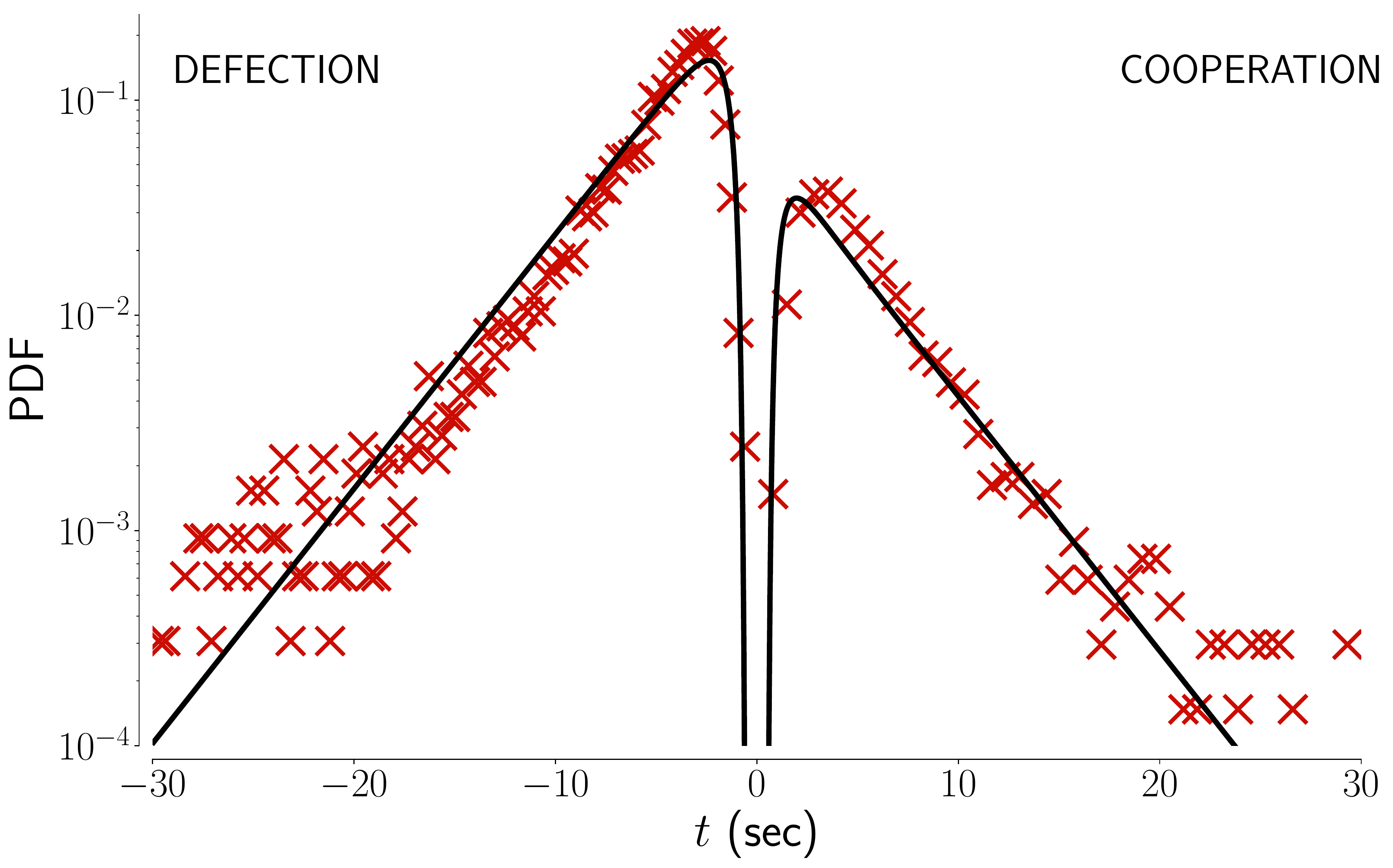}&
\raisebox{2.5cm}{(b)} \includegraphics[angle=0, width=0.45\textwidth]{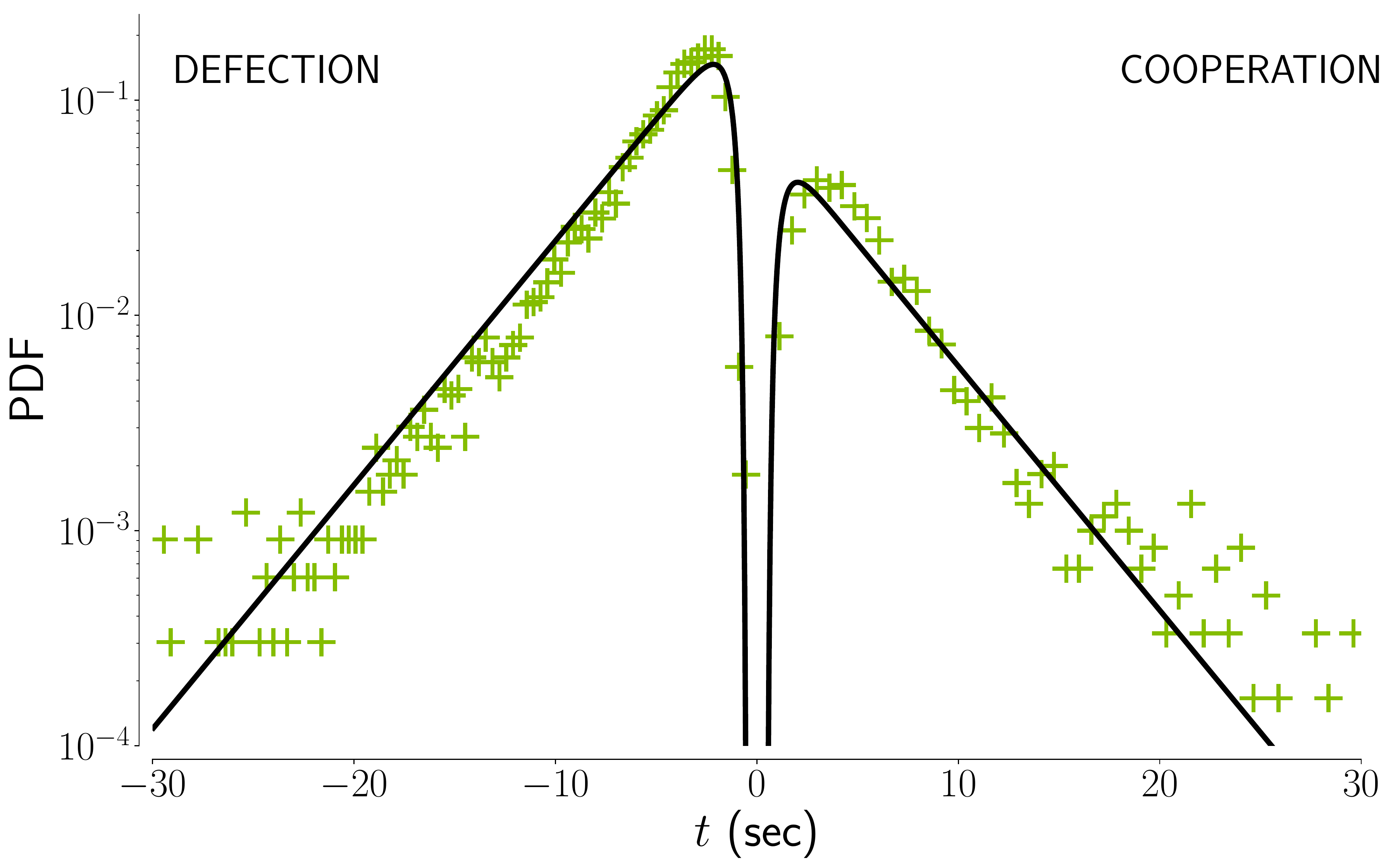}
\end{tabular}
\begin{center}
\raisebox{2.5cm}{(c)} \includegraphics[angle=0, width=0.45\textwidth]{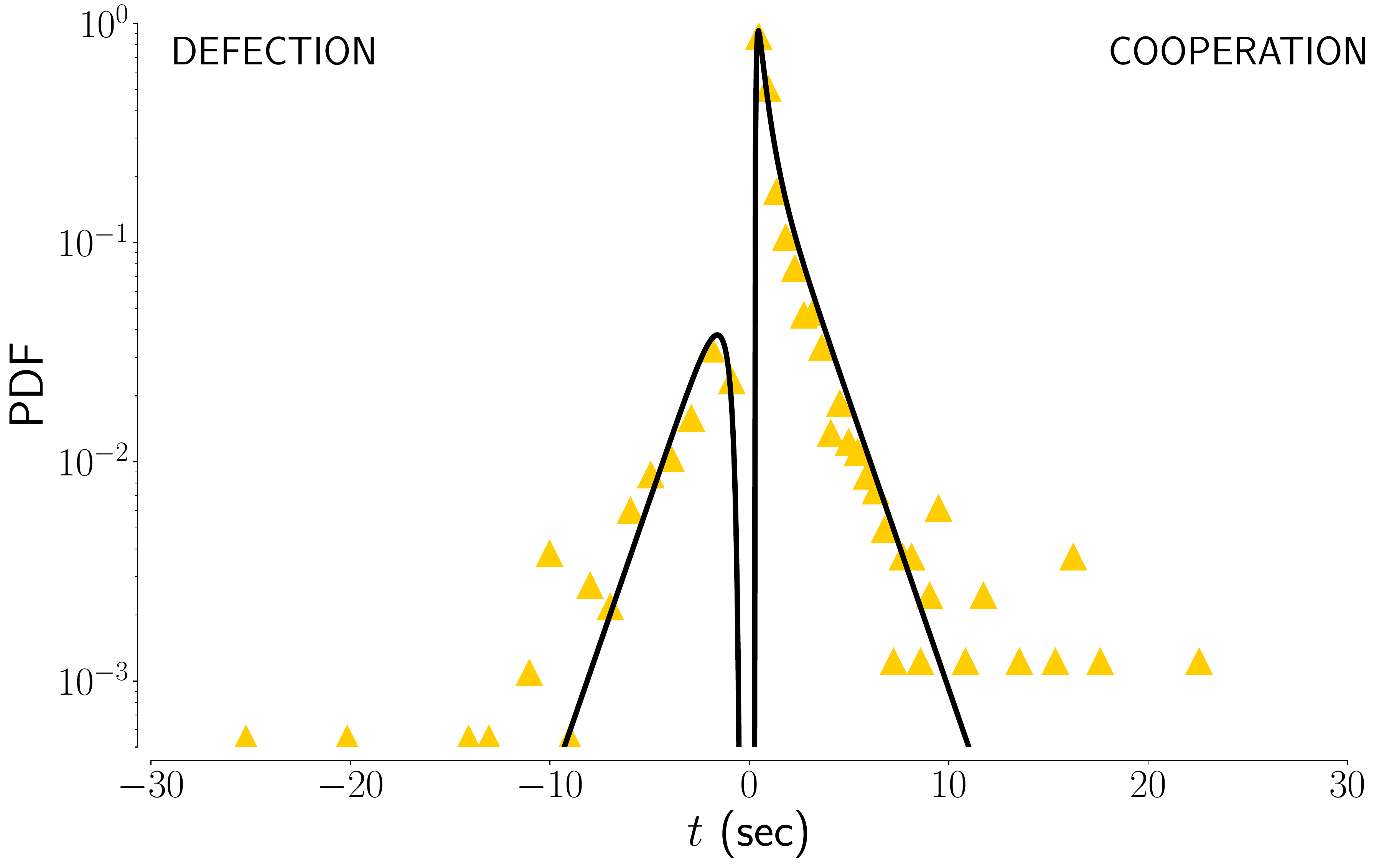}
\end{center}
\caption{{\bf Experimental distribution fitted with the theoretical curves for the DDM.} 
{\bf (a)} Multiplayer experiment, random phase  ($r^2 = 0.94$).
{\bf (b)} Multiplayer experiment, fix 2 phase  ($r^2 = 0.95$).
{\bf (c)} Pairwise experiment  ($r^2 = 0.99$).
}
\label{SI:figure3}
\end{figure*}


\begin{figure*}[ht!]
\def\figurename{Supplementary Figure}
\centerline{
\includegraphics[angle=0,width=0.48\textwidth]{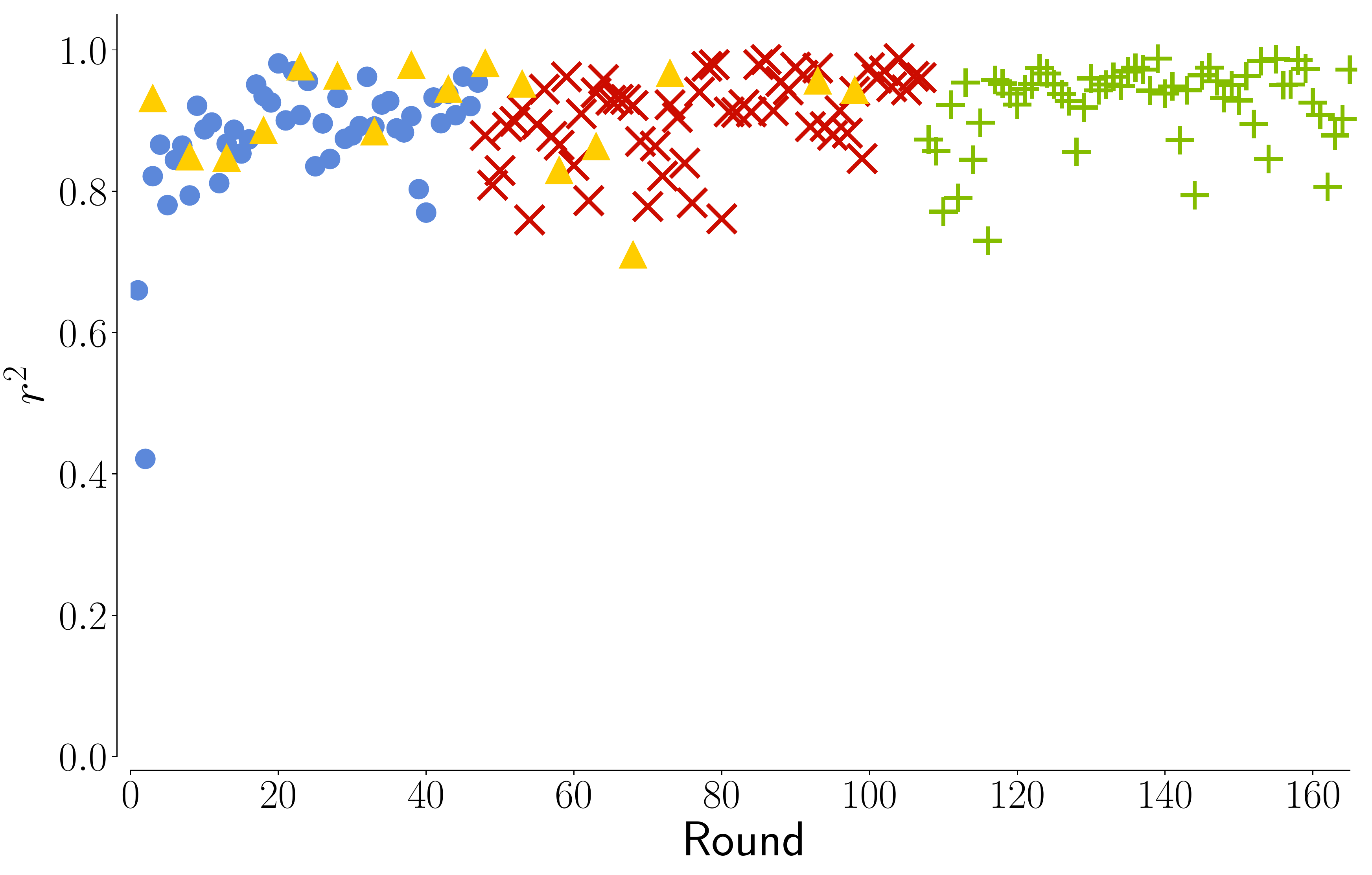}
}
\caption{{\bf $r^2$} value of the fits described in Figure 3.}
\label{SI:figure4}
\end{figure*}

\begin{figure*}[ht!]
\def\figurename{Supplementary Figure}
\centerline{
\includegraphics[angle=0,width=0.8\textwidth]{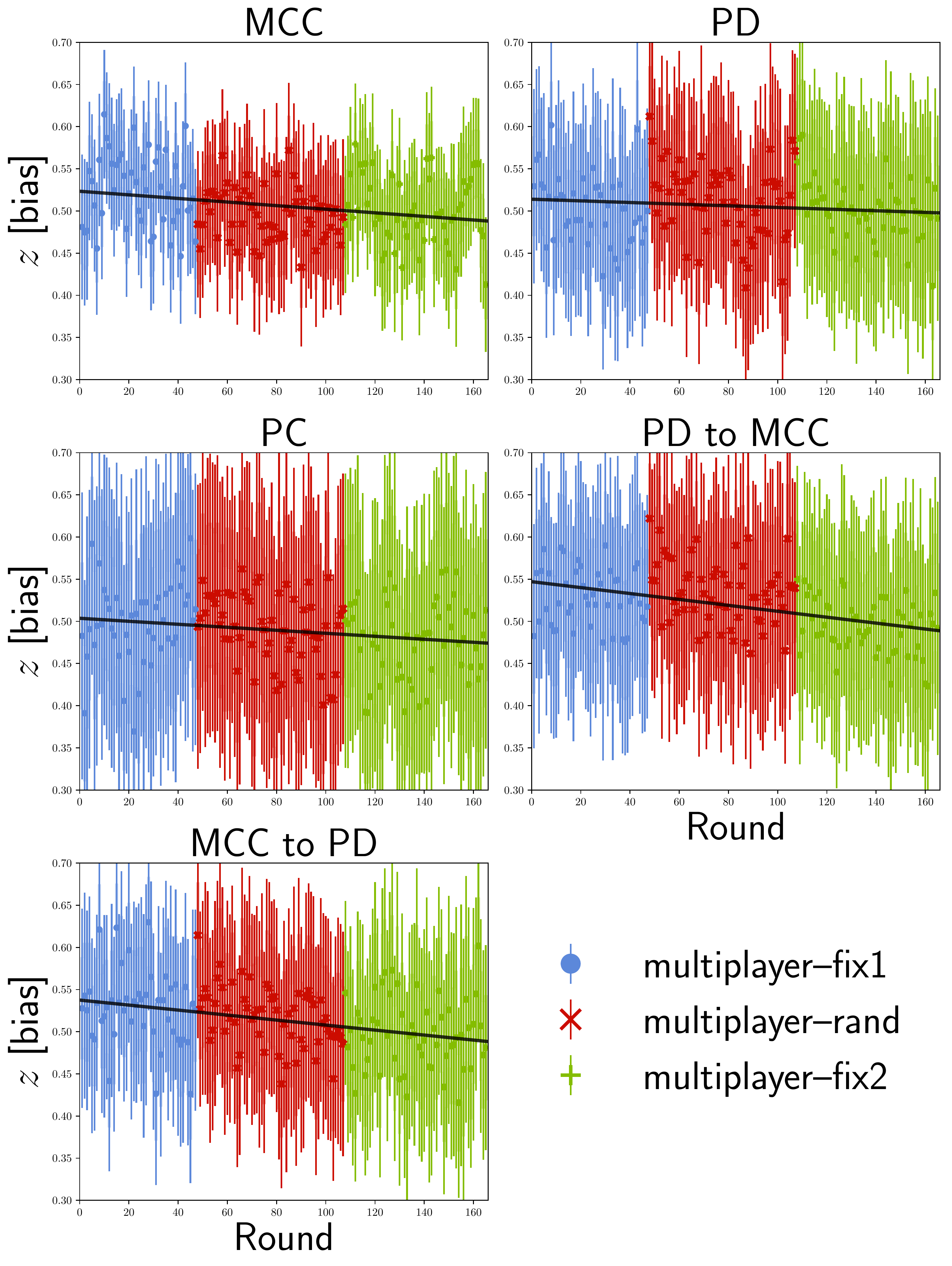}
}
\caption{
{\bf The drop in the value of the bias $z$ is independent by the players' strategy.} 
In the multiplayer experiment we can separate players into three types: Moody Conditional Cooperators (MCC), Pure Defectors (PD), and Pure Cooperators (PC)
[20]. 
PD and PC are the players who always defect or cooperate 
and MCC change their action depending on what their neighbours did in the previous round and what they themselves did in the previous round. Here we separate players by the strategy they picked in fix1 and fix2: in the first three panels we show the results for users who kept the same strategy in both phases, while in the latter two users who switched between two different strategies. In all the cases we observe a similar average drop in the value of the bias $z$. The progressive emergence of a bias towards defection is therefore not related to the users actions or strategy. Consequently, we argue that it necessarily depends on the experience of playing against a majority of defectors.}
\label{SI:figure5}
\end{figure*}


\begin{table*}[h]
\def\tablename{Supplementary Table}
\begin{center}
\ra{1.3}
\begin{tabular*}{\textwidth}{@{\extracolsep{\fill}}lcccc@{}}
\toprule[1pt]
&multiplayer fix1 & multiplayer rand & multiplayer fix2 & pairwise\\
\hline
Rounds				& 47					& 60					& 58					& 100			\\
$\langle t \rangle$ (sec)	& $6.73$				& $5.02$				& $4.98$				& $1.59$			\\
STD$(t)$ (sec)			& $4.56$				& $3.55$				& $3.57$				& $2.06$			\\
SEM$(t)$ (sec)			& $0.67$				& $0.46$				& $0.47$				& $0.21$			\\
$C_R$				& $28.7\%$			& $18.5\%$			& $22.9\%$			& $88.8\%$		\\
\hline
$a$ (sec$^{\frac{1}{2}}$)	& $5.32 \pm 0.03$  		& $4.86 \pm 0.02$ 		& $4.73 \pm 0.02$		& $3.01 \pm 0.04$	\\
$v$ (sec$^{-\frac{1}{2}}$)	& $-0.203 \pm 0.005$	& $-0.357 \pm 0.006$ 	& $-0.284 \pm 0.006$	& $0.35 \pm 0.02$	\\
$z$ 					&$0.529 \pm 0.004$ 		& $0.529 \pm 0.004$ 	& $0.509 \pm 0.004$ 	& $0.731 \pm 0.006$\\ 
$t_0$ (sec)			& $0.38 \pm 0.01$ 		& $0.340 \pm 0.007$ 	& $0.315 \pm 0.005$		& $0.236 \pm 0.003$	\\
$r^2$ 				& $0.97$ 				& $0.94$ 				& $0.95$				& $0.99$			\\
\bottomrule
\end{tabular*}
\end{center}
\caption{
{\bf Average and fit values for the whole phases of the multiplayer experiment and for the pairwise experiment} 
\textnormal{ These values describe the distribution and fits in Figure 2 and Supplementary Fig. 3. We remark that between the phases fix2 and rand one would not see any significative differences if the study were considering only the moments of the decision times distribution. }
}
\label{SI:Table1}
\end{table*} 

\end{document}